\documentclass[journal,twoside]{IEEEtran}

\IEEEoverridecommandlockouts

\usepackage{cite}

\usepackage{textcomp}

\usepackage{multirow}
\usepackage[table,xcdraw]{xcolor}
\usepackage{tabularx} 
\usepackage{tcolorbox} 
\usepackage{amsfonts}
\usepackage{mathrsfs}
\usepackage{algorithm}          
\usepackage{algpseudocode}     
\usepackage{amsmath}            
\usepackage{graphicx}

\usepackage{subcaption} 
\usepackage{enumitem}
\usepackage{xurl}
\usepackage{amssymb}  
\usepackage[colorlinks=false, allcolors=blue, pdfborder={0 0 0}]{hyperref}

\newcolumntype{Y}{>{\centering\arraybackslash}X}
\newcolumntype{H}{>{\columncolor[gray]{0.9}\centering\arraybackslash}X}

\begin{document}

\title{Breaking the Black Box: Byte-Level Boundary Inference of Real-World Antivirus Systems}

\author{Jieshuai~Yang,
    Zhi~Wang,
    Yan~Jia,
    Zhenhua~Wu,
    Jianfei~Tang,
    Chenbin~Su,
    Jingwei~Ye,
    Jianwen~Tian,
    and Wanpeng~Li%
    \thanks{(Corresponding author: Zhi~Wang.)}%
    \thanks{Jieshuai~Yang, Zhi~Wang, Yan~Jia, Zhenhua~Wu,
    Jianfei~Tang, Chenbin~Su, and Jingwei~Ye are with the
    College of Cryptology and Cyber Science, Nankai University,
    Tianjin 300350, China
    (e-mail: yjs@mail.nankai.edu.cn;
    zwang@nankai.edu.cn;
    jiay@nankai.edu.cn;
    2120250720@mail.nankai.edu.cn;
    kid519388073@mail.nankai.edu.cn;
    champion.su@mail.nankai.edu.cn;
    jwye@mail.nankai.edu.cn).}%
    \thanks{Jianwen~Tian is with the School of Computing and
    Information Systems, Singapore Management University,
    80 Stamford Road, Singapore 178902
    (e-mail: jwtian@smu.edu.sg).}%
    \thanks{Wanpeng~Li is with the School of Computer Science
    and Informatics, University of Liverpool, Liverpool, UK
    (e-mail: wanpeng.li@liverpool.ac.uk).}%

}


\markboth{}
{Yang \MakeLowercase{\textit{et al.}}: Breaking the Black Box: Byte-Level Boundary Inference of Real-World Antivirus Systems}


\maketitle

\begin{abstract}
Existing approaches for understanding the detection logic of real-world antivirus (AV) software infer only binary malware/benign decisions from black-box queries, providing limited insight into the fine-grained decision-critical regions that govern AV detection. In this paper, we present \textbf{AVHunter}, the first framework for inferring byte-level decision-critical regions of real-world AV products under a black-box threat model. AVHunter constructs the first large-scale Byte-Level AV Boundary Dataset (BABD) by systematically probing 11 real-world AV products, revealing that modern AV detections are largely associated with a small number of compact decision-critical byte regions. Leveraging BABD, AVHunter trains AV-specific models that not only reproduce binary AV decisions, but also localize the decision-critical byte regions underlying these decisions, achieving an average boundary prediction recall of 85.07\% while maintaining 97.43\% detection agreement with the target AVs. We further validate that the predicted regions capture genuine AV decision knowledge through boundary-guided malware evasion, false-positive induction on benign executables, and a seven-month longitudinal study demonstrating that the inferred regions remain largely stable as AV products evolve. Overall, AVHunter moves beyond conventional binary-label AV modeling by enabling fine-grained boundary-region localization and revealing a new form of AV knowledge leakage with important implications for malware analysis, AV security, and boundary-aware defenses.
\end{abstract}

\section{Introduction}

Model-stealing techniques have become a well-established threat in both the computer vision~\cite{chen2023d,sha2023can,zhao2024fully,zhuang2025stealix,luan2025dynamic,yuan2024data} and natural language processing domains~\cite{ICLR2025_cce0e917,karmakar2023marich,liu2025model,naseh2023stealing}. By enabling the construction of high-fidelity surrogate models, these attacks can leak proprietary intellectual property and facilitate effective adversarial sample generation, leading to substantial economic and security risks for model vendors.

Prior studies have shown that model-stealing attacks can extend to real-world antivirus (AV) software, where black-box queries and semi-supervised active learning can be used to construct surrogate models with high detection agreement~\cite{rigaki2023stealing}. Malware evasion studies similarly rely on surrogate models or optimization objectives that implicitly learn decision boundaries to guide functionality-preserving modifications~\cite{demetrio2021functionality,ling2024wolf}. However, these learned boundaries are typically derived from coarse-grained features or substitute detectors, and thus provide limited effectiveness against real-world AV products. As a result, high surrogate agreement or successful surrogate-side evasion does not necessarily indicate accurate recovery of the AV's fine-grained byte-level boundary.

We attribute this limitation primarily to the discrete nature of AV decision boundaries and the strict black-box constraints under which AV engines operate. Although machine learning-based malware detectors have shown promising results in controlled settings, their practical deployment is often hindered by high false-positive rates and substantial maintenance costs~\cite{cavallaro2023machine}. In practice, real-world AV products are hybrid systems that combine multiple detection sources, such as signatures, cloud intelligence, behavioral profiles, and runtime monitoring. However, prior empirical analysis of commercial AV internals shows that most detection capabilities are still provided by static checks~\cite{botacin2022antiviruses}. Consequently, many AV decisions remain closely tied to discrete static detection patterns, where specific combinations of key bytes can determine whether a sample matches a detection rule~\cite{botacin2022antiviruses,wressnegger2017automatically}. Existing model-stealing approaches, however, only observe binary benign/malware feedback and cannot access the internal detection logic of AV engines. Such coarse supervision loses fine-grained boundary information, causing surrogate models to approximate global AV behavior while failing to accurately reconstruct the byte-level decision-critical regions near the true AV boundary.
To construct a surrogate model that can both replicate AV behavior and recover actionable byte-level boundary knowledge, two key challenges must be addressed:

\textbf{C1: Extracting and representing AV decision boundary information in a learnable form under black-box constraints.}
Because AV engines expose only binary detection feedback, existing stealing approaches suffer from severe information loss. The challenge is to identify the sparse yet critical byte patterns that govern AV detection and transform such boundary information into machine-learnable supervision without accessing internal signatures, rules, or models.

\textbf{C2: Learning extremely sparse boundary signals within large binary inputs.}
AV boundary evidence is sparse and embedded in high-dimensional binary inputs. In a 1 MB executable, most bytes are irrelevant to detection, while the decision-critical bytes, which we define as boundary points, are typically fewer than 20 bytes (see Section~\ref{Section:Decision boundary}). Modifying these bytes can invalidate the corresponding detection pattern and cause the AV to miss the malware. Therefore, the surrogate model must reliably isolate weak boundary signals and learn them precisely, rather than producing noisy or overly coarse boundary estimates.

To address these challenges, we present \textbf{AVHunter}, a novel framework for systematically inferring fine-grained decision boundaries of real-world AV products at the byte level. \textbf{To address C1}, AVHunter performs controlled perturbations on malware samples to identify decision-critical bytes and expands them into boundary regions that preserve contextual information, thereby converting localized AV boundary evidence into learnable byte-level labels. \textbf{To address C2}, we design AV-DBNet, a boundary-aware surrogate model based on a multi-scale convolutional encoder, attention fusion, and decoder architecture. AV-DBNet jointly learns AV classification behavior and boundary probability distributions, enabling it to approximate target AV decisions while explicitly recovering decision-critical byte regions for downstream tasks such as boundary-guided evasion analysis, AV detection logic analysis, and cross-AV knowledge transfer assessment.

Using AVHunter, we construct the Byte-Level AV Boundary Dataset (BABD), a large-scale dataset comprising \texttt{241,070} malware and benign executables annotated with byte-level boundary information from 11 real-world AV products. BABD reveals that, despite the increasing complexity of modern AV systems, their detections are largely associated with a small number of compact decision-critical byte regions. Leveraging BABD, we train AV-specific models that not only reproduce binary AV decisions, but also predict the decision-critical byte regions underlying these decisions. Extensive experiments across 11 real-world AV products demonstrate that AVHunter achieves 85.07\% average boundary prediction recall while maintaining 97.43\% detection agreement with the target AVs. The predicted regions further enable highly effective boundary-guided malware evasion, achieving an average evasion rate of 83.78\%. In a seven-month longitudinal study, AVHunter remains largely stable, with detection agreement decreasing by only 1.59\%.

In summary, this paper makes the following contributions:

\begin{itemize}
\item We construct \textbf{BABD}, the first large-scale byte-level antivirus boundary dataset spanning 11 real-world AV products. BABD provides a benchmark for AV boundary-region localization and reveals that AV detections rely on a small number of compact decision-critical byte regions.
\item We propose \textbf{AVHunter}, the first black-box framework for localizing byte-level decision-critical regions of real-world antivirus products. Unlike prior AV model-stealing approaches that reproduce only binary malware/benign labels, AVHunter predicts fine-grained decision-critical byte regions while maintaining high agreement with target AV detection.
\item We perform the first large-scale evaluation of AV boundary-region localization across 11 real-world AV products. The results demonstrate that AVHunter accurately predicts decision-critical regions, which are validated through boundary-guided malware evasion, false-positive induction, temporal robustness, and comparisons with state-of-the-art binary-feedback-only approaches.
\item We show that AV knowledge leakage goes beyond binary-label imitation. Fine-grained boundary knowledge can be recovered from black-box interactions, exposing risks of unauthorized AV replication, boundary-aware malware attacks, and transferable malware analysis.

\end{itemize}

Our implementation and open-science policy are available at \url{https://anonymous.4open.science/r/AVHunter-8BE2/}.

\section{Preliminaries and Threat Model}

\subsection{Preliminaries}

\subsubsection{Decision boundary}

\label{Section:Decision boundary}
\begin{figure}[t]\centering
	\includegraphics[width=0.8\linewidth]{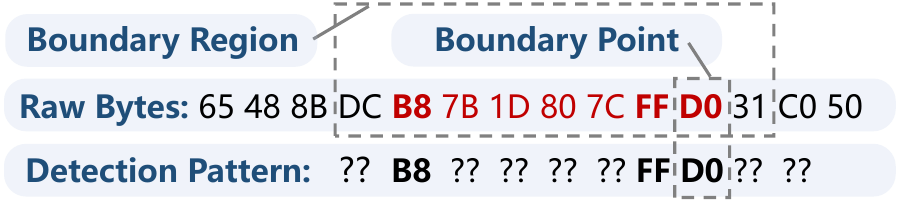}
	\caption{Illustration of AV Detection Patterns, Boundary Points, and Boundary Regions. Each detection pattern (e.g., signature or rule) contains critical Boundary Points (BPs), which, if modified, can invalidate the pattern. The surrounding context forms a Boundary Region (BR); modifying a BR overwrites the enclosed BPs and disrupts the associated detection rule.}
	\label{fig:Comparison of Signature Rule Pattern BP and BR}
\end{figure}

In this paper, we define the decision boundary of an AV engine as the set of discrete patterns, such as signatures or rules, which determine whether a sample is flagged as malicious. The critical byte positions within these patterns are referred to as Boundary Points (BPs); modifying these bytes invalidates the corresponding pattern match. The surrounding contexts of BPs are termed Boundary Regions (BRs). Rewriting a BR effectively overwrites the enclosed BPs, thereby disrupting the associated detection pattern. Figure~\ref{fig:Comparison of Signature Rule Pattern BP and BR} illustrates the relationships among these concepts. 

It is important to note that our definition of rule patterns is broad. Modern AVs typically employ hybrid detection strategies, and components such as ML-based detection and statistical heuristics may introduce a large number of BPs, making precise inference considerably more difficult. Nevertheless, statistical examination of BABD indicates that, despite this potential complexity, boundary-point cardinality remains limited in the overwhelming majority of instances. Concretely, 93.26\% of evaluated samples across the 11 AV engines exhibit fewer than 20 BPs (see Section~\ref{Section:Probing Scope}). This observation forms the empirical basis for pursuing fine-grained, byte-level boundary inference in black-box settings.

\begin{figure*}[t]\centering
	\centering
	\includegraphics[width=0.8\linewidth]{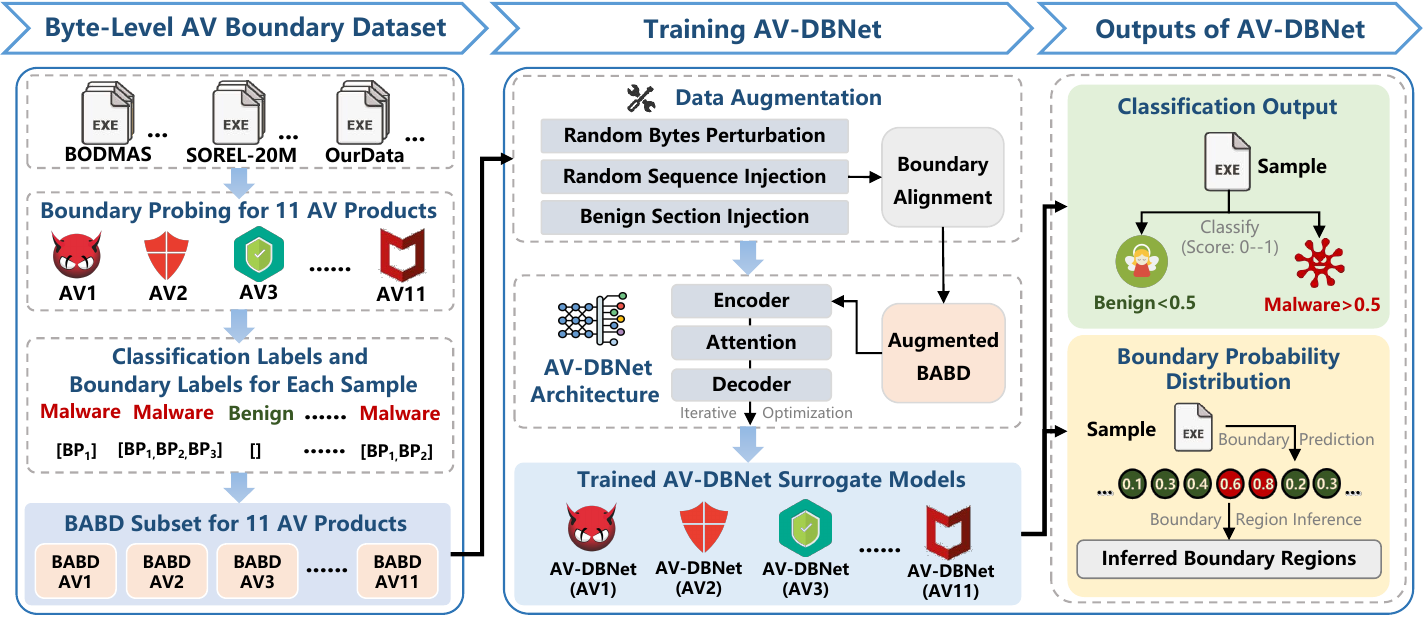}
	\caption{{ The AVHunter framework comprises two core stages: probing malware to extract BPs and construct the Byte-Level AV Boundary Dataset, which captures byte-level AV decision boundary information for 11 target AV products; and training the multi-scale encoder--attention fusion--decoder surrogate model AV-DBNet for each AV product on the augmented data. For each input sample, the trained AV-DBNet outputs two results: a binary classification score and a boundary probability distribution.}}

	\label{fig:AVHunter Framework}
\end{figure*}

\subsubsection{Boundary Inconsistency}
\label{Section:Boundary Inconsistency}

Prior AV model-stealing studies~\cite{rigaki2023stealing} can achieve high agreement with real-world AV products, while surrogate-guided evasion methods~\cite{demetrio2021functionality,ling2024wolf} have shown effective evasion against ML-based detectors. However, these approaches often fail to capture the AV's fine-grained decision boundary. This limitation becomes evident under boundary-sensitive modifications: modifying decision-critical bytes may disrupt the AV detection pattern and cause the AV to miss the sample, while a coarse-grained surrogate may still classify it as malicious due to only small feature-space shifts. Conversely, injection-based attacks~\cite{demetrio2021functionality,zapzalka2024semantics,ling2024wolf,zhang2023semantics} may mislead the surrogate with benign-looking features, while the AV still detects the sample if its original detection patterns remain intact. These cases show that existing surrogate-based approaches mainly approximate global decision behavior rather than the true decision-critical regions used by AV products.

Therefore, an effective AV surrogate should recover BRs covering decision-critical bytes. Our boundary-guided malware modification study supports this requirement: generic evasion methods without boundary localization struggle to bypass target AVs, whereas AVHunter uses inferred BRs to obtain precise boundary knowledge and achieve substantially higher evasion rates, as shown in Section~\ref{Section:Boundary-Guided Malware Modification}. Such boundary information helps the surrogate better approximate AV decision boundaries and achieve high detection agreement (see Section~\ref{section:Detection Agreement}), while remaining more consistent under boundary-sensitive perturbations (see Section~\ref{Section:Detection Resilience Against Adversarial Attacks}).

\subsection{Threat Model}

\textbf{Assumptions.}
We consider Windows-based real-world AV engines in a black-box setting and focus on their static detection interfaces, where the AV returns a binary hard-label detection outcome (malicious or benign) for a given executable file without exposing internal signatures, models, or confidence scores. Our analysis focuses on malware samples whose detection can be associated with a limited number of BPs, which account for 93.26\% of evaluated samples on average across the 11 AV products (see Section~\ref{Section:Probing Scope}).

\textbf{Adversary Capabilities.}
The adversary can repeatedly query the target AV with crafted binaries and observe detection results, but has no access to internal signatures, models, or confidence scores. The attacker can generate modifications to malware binaries and collect query–response pairs for surrogate training.

\textbf{Adversary Goals.}
The adversary’s objective is to infer the target AV’s fine-grained decision boundary by constructing a high-fidelity surrogate model. Formally, given black-box access to an AV with detection function $f_{AV}$, the adversary seeks to learn a surrogate $f_{sub}$ that closely approximates $f_{AV}$ over relevant inputs. Accurate boundary inference enables recovery of decision-critical boundary knowledge, which can support downstream tasks, such as boundary-guided evasion analysis, AV detection logic analysis, and cross-AV knowledge transfer assessment.

\section{Approach}

As illustrated in Figure~\ref{fig:AVHunter Framework}, \textbf{AVHunter} follows a two-stage pipeline. First, it systematically probes malware samples to identify BPs and constructs a dataset that captures byte-level AV decision boundary information for 11 target AV products. Second, for each AV product, AVHunter trains a multi-scale convolutional encoder-attention fusion-decoder model, termed AV-DBNet, on the corresponding dataset to learn fine-grained AV-specific decision boundaries. For each input sample, the trained AV-DBNet produces two outputs: a binary classification score and a boundary probability distribution.

\begin{algorithm}[!t]
\caption{Pseudocode of Boundary Probing and Labeling}
\label{alg:Pseudocode of Boundary Probing and Labeling}
\begin{algorithmic}[1]

\Require Malware sample set $M$, target AV $f_{AV}$, params $k, h$
\Ensure Boundary label $\mathcal{Y}_{bound}$, classification label $\mathcal{Y}_{class}$

\For{each $m \in M$}
    \State Initialize $m_n \leftarrow m$, $\mathcal{B}_m \leftarrow \emptyset$
    \While{$f_{AV}(m_n) \neq 0$}
        \State Initialize $\text{low} \gets 0$, $\text{high} \gets Len(m_n)$
        \While{$\text{high} - \text{low} > 1$}
            \State $\text{mid} \gets \lfloor(\text{low} + \text{high})/2\rfloor$
            \State $m_t \gets \text{Mask the range } [\text{mid}, Len(m_n)] \text{ in } m_n$
            \State Query $f_{AV}(m_t)$
            \If{$f_{AV}(m_t) = 0$}
                \State $\text{low} \gets \text{mid}$
            \Else
                \State $\text{high} \gets \text{mid}$
            \EndIf
        \EndWhile
        \State $b_p \gets \text{low}$, $\mathcal{B}_m \gets \mathcal{B}_m \cup \{b_p\}$
        \State $m_n \gets \text{Mask all BPs in } \mathcal{B}_m \text{ from } m$
    \EndWhile
\EndFor

\State $\mathcal{Y}_{bound} \gets$ for each $m$, set 1s at indices $[b_p-k, b_p+h]$ for $b_p \in \mathcal{B}_m$, and 0s elsewhere
\State $\mathcal{Y}_{class} \gets 1$ for malware samples, $0$ for benign samples

\end{algorithmic}
\end{algorithm}

\subsection{Byte-Level AV Boundary Dataset}
\label{Section:Byte-Level AV Boundary Dataset}

AVHunter constructs a \emph{Byte-Level AV Boundary Dataset (BABD)} by probing malware samples to recover fine-grained byte-level decision boundaries from target AV engines. For each sample, boundary annotations are extracted from all 11 AV products, producing AV-specific boundary labels. 

\subsubsection{Probing Procedure}

Our probing process is inspired by prior black-box detector testing and AV signature-inference studies, which show that controlled binary modifications and AV feedback can expose detection-sensitive byte patterns~\cite{christodorescu2004testing,wressnegger2017automatically,blackthorne2016avleak}. AVHunter uses an efficient iterative method to probe multiple BPs per sample and construct AV-specific byte-level boundary annotations, as summarized in Algorithm~\ref{alg:Pseudocode of Boundary Probing and Labeling}.

\textbf{(1) Boundary Point Probing.}
Given a malware sample detected by an AV, AVHunter identifies its BPs through iterative binary search. Starting from the whole file, it repeatedly masks the suffix region $[\textit{mid}, \textit{Len}(m)]$ with the neutral byte \texttt{0x90} and submits the modified sample to the target AV. This tail-oriented masking preserves the PE header and critical metadata in the prefix region, reducing false benign outcomes caused by structural corruption. If the target byte is already \texttt{0x90}, AVHunter replaces it with a random non-\texttt{0x90} byte, excluding \texttt{0x00} to avoid disrupting PE parsing. When the AV classifies the masked sample as benign, i.e., $f_{\text{AV}}(m_t)=0$, the masked suffix is considered to contain decision-critical malicious patterns; otherwise, the search continues in the unmodified region. The process stops when the interval converges to a single byte, which is recorded as a boundary point $b_p$.

\textbf{(2) Multiple Boundary Point Handling.}
Malware samples may contain multiple non-contiguous BPs corresponding to different detection patterns. After finding one BP, AVHunter masks it with \texttt{0x90} to obtain a new sample $m_n$ and repeats the probing process until no additional BP can be found, i.e., $f_{\text{AV}}(m_n)=0$. Adjacent BPs are merged into compact multi-byte annotations. A small number of BPs may correspond to critical PE attributes, such as header flags; we treat them as valid boundary rules because they still reveal byte-level conditions affecting AV decisions.

\textbf{(3) Boundary Region Construction.}
To preserve contextual information, AVHunter expands each BP into a boundary region (BR). For a BP $b_p$, the BR is defined as $[b_p-k, b_p+h]$, where $k$ and $h$ denote the preceding and following context lengths. We set $k=30$ and $h=10$, since most observed AV patterns fall within this range. The larger preceding context is used because probing often locates the last byte of a pattern, and backward expansion helps recover the complete pattern context. The final annotation is $\mathcal{Y}_{\text{bound}}=\bigcup_{i=1}^{N} I[b_p^{(i)}-k,b_p^{(i)}+h]$, where $b_p^{(i)}$ is the $i$-th BP. This BR representation localizes AV decision boundaries while retaining context for downstream surrogate learning.

\begin{table}[t]

\renewcommand{\arraystretch}{1.3}

\caption{{Composition of the Byte-Level AV Boundary Dataset (BABD). The dataset contains malware and benign samples with byte-level boundary labels from 11 real-world AV engines, split into training and testing subsets. The ``$\times 11$'' notation denotes AV-specific labeled instances, since each sample is annotated separately for 11 AV products.}}
	\label{tab:Dataset Setting}
    \centering
    \resizebox{1\linewidth}{!}{\large 
\begin{tabular}{ccccccc}
\hline
\multirow{2}{*}{\textbf{Dataset}} & \multicolumn{2}{c}{\textbf{Training Set}} &  & \multicolumn{2}{c}{\textbf{Testing Set}} & \multirow{2}{*}{\textbf{Total}} \\ \cline{2-3} \cline{5-6}
                                  & \textbf{Malware}    & \textbf{Benign}    &  & \textbf{Malware}    & \textbf{Benign}    &                                 \\ \hline
BODMAS                            & 25000               & -                  &  & 5000                & -                  & 30000                           \\
SOREL-20M                         & 50341               & -                  &  & 10078               & -                  & 60419                           \\
OurData                           & 25137               & 100413             &  & 5017                & 20084              & 150651                          \\
Total                             & 100478×11           & 100413×11          &  & 20095×11            & 20084×11           & 241070×11                       \\ \hline
\end{tabular}
}
\end{table}

\begin{figure}[t]
\centering
\captionsetup{skip=1pt}
\includegraphics[width=\linewidth]{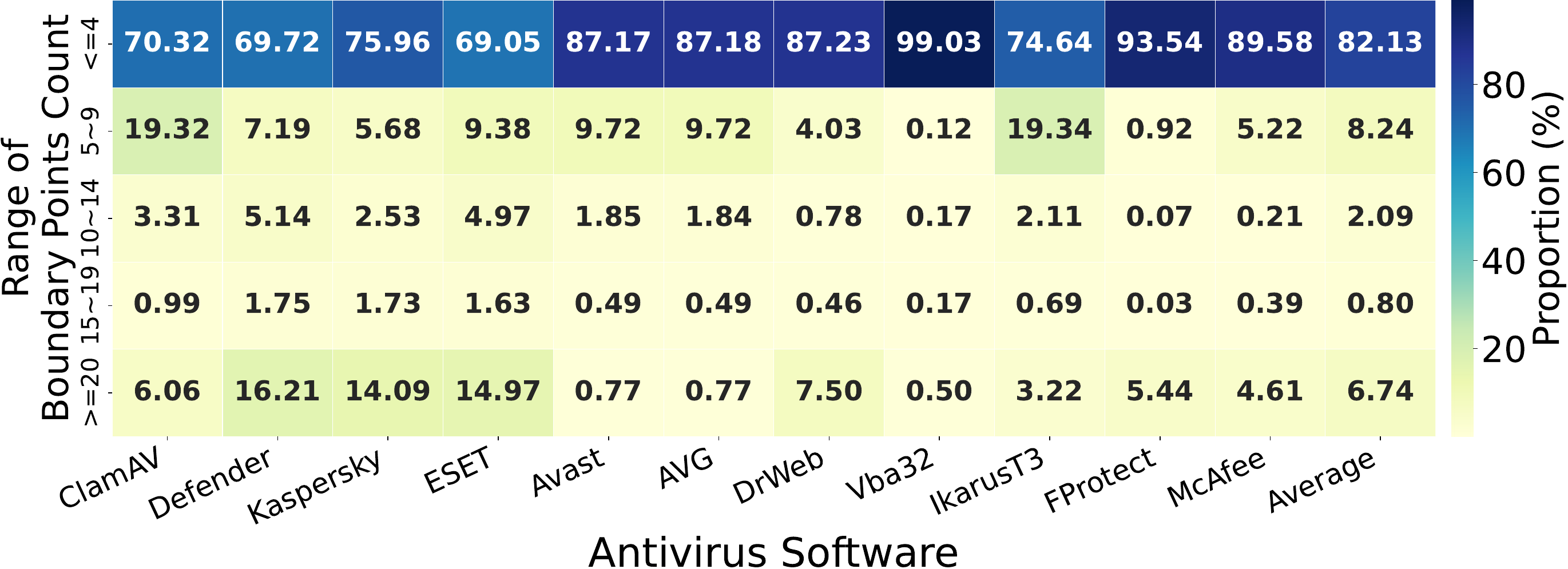}
\caption{Heatmap of the distribution of boundary-point counts in the BABD dataset across malware samples and AV products.}
\label{fig:Boundary Points Distribution Heatmap}
\end{figure}
\subsubsection{Sample Collection}
Malware samples in BABD are collected from BODMAS~\cite{yang2021bodmas}, SOREL-20M~\cite{harang2020sorel}, and an internal corpus derived from VirusShare~\cite{virusshare} and VirusTotal~\cite{virustotal}, covering 605 families according to Windows Defender labels. Since public datasets rarely provide benign executables due to copyright restrictions, benign samples are collected from widely used commercial PE software and Windows system executables. To control probing cost, all samples are capped at 1 MB and analyzed against 11 AV products (see Section~\ref{Section:4.1.2}). Probing requires 66.53 queries per sample, 19.20 queries per BP, and 1.91 seconds per sample on average. BABD contains \texttt{241,070} samples with \texttt{2,651,770} byte-level boundary annotations. Dataset statistics are summarized in Table~\ref{tab:Dataset Setting}, and the 11 AV-specific BABD subsets are used to train and evaluate the corresponding AV-DBNet models.

\subsubsection{Boundary-Point Cardinality Analysis}
\label{Section:Probing Scope}

We acknowledge that modern AV engines may use heterogeneous detection mechanisms, such as statistical heuristics or ML-based classifiers, which could produce many BPs and make exhaustive probing impractical. To quantify this issue, Figure~\ref{fig:Boundary Points Distribution Heatmap} visualizes the BP-count distribution of malware samples across the 11 AV engines in BABD. The results show that 93.26\% of evaluated samples contain fewer than 20 BPs, indicating that most static AV detections can be associated with a limited number of decision-critical bytes. Therefore, AVHunter imposes a practical upper limit of 20 BPs per sample, ensuring computational feasibility while still covering the dominant detection behaviors observed in real-world AV products. 

\subsubsection{Ground-Truth Validation}
To verify whether probing identifies bytes tied to real detection mechanisms, we build ground-truth detection regions from the open-source ClamAV~\cite{ClamAV2026} signature database. We obtain the triggered ClamAV signatures for each malware sample from BABD, parse their content and locatable hash signatures, recover the matching regions in the original PE files, and compare them with AVHunter’s probing bytes. Ground-truth regions can be recovered for 96.64\% of samples; among them, 97.39\% show overlap, indicating that AVHunter’s critical bytes are consistent with the actual evidence used by antivirus engines. The remaining 2.61\% may result from multiple triggered signatures, compound detection logic, or probing locating related contextual bytes rather than the exact signature region.


\subsection{Training AV-DBNet}

To model fine-grained AV decision boundaries, we design AV-DBNet, a multi-scale convolutional encoder–attention fusion–decoder architecture. Unlike models that process only local features, AV-DBNet leverages multi-scale convolutions to extract hierarchical features at multiple granularities, enabling accurate boundary prediction while maintaining holistic malware classification. For each of the 11 AV products, AV-DBNet is trained on a dedicated subset of BABD (see Section~\ref{Section:Byte-Level AV Boundary Dataset}), which provides byte-level annotations of that AV’s decision boundaries. This targeted training allows the model to learn critical boundary positions and surrounding contextual patterns specific to each AV’s detection logic.


    


\subsubsection{Data Augmentation and Boundary Alignment}

To improve robustness against adversarial evasion, AV-DBNet is trained with perturbation-augmented samples, including random byte perturbation, random sequence injection, and benign section injection, which emulate byte-level changes, layout shifts, and benign feature injection. AVHunter also aligns boundary labels after augmentation: random byte perturbation and benign section injection preserve $\mathcal{Y}_{\text{bound}}$, while random sequence injection shifts byte offsets and requires displacement-based alignment to produce $\mathcal{Y}_{\text{bound}}'$.

\subsubsection{Model Architecture}

AV-DBNet adopts a multi-scale encoder--attention fusion--decoder architecture for byte-level boundary inference. It consists of three components: an encoder for hierarchical feature extraction, an attention module for feature refinement, and a decoder with dual heads for boundary prediction and classification.

\textbf{(1) Encoder Module.}
The encoder embeds the input byte sequence and adds learnable positional encodings~\cite{devlin2019bert} to capture sequential information. The resulting representation is processed by multi-scale convolutions to extract hierarchical contextual features at different granularities, which supports accurate boundary point prediction.

\textbf{(2) Attention Fusion Module.}
This module applies the dual-attention mechanism from~\cite{woo2018cbam}, which combines channel and spatial attention to enhance discriminative features. By adaptively weighting important channels and byte regions, it helps the model focus on boundary-critical patterns.

\textbf{(3) Decoder Module.}
The decoder uses a dual-output design. The boundary head upsamples features through transposed convolutions and produces the boundary probability distribution $\hat{\mathcal{Y}}_{\text{bound}}$. To complement local boundary knowledge with broader sample context, the classification head further introduces a global raw-byte feature branch and fuses it with predicted BR-based regional features to generate the classification score $\hat{\mathcal{Y}}{\text{class}}$.

\subsubsection{Loss Function}
AV-DBNet employs a composite loss function consisting of three key components:

\begin{itemize}
    \item \textbf{Focal loss}~\cite{lin2017focal} addresses the extreme class imbalance in boundary prediction 
    (positive bytes $<1\%$) through a dynamic weighting mechanism
    $\mathcal{L}_f = -\sum \alpha_t (1-q_t)^\gamma 
    \mathcal{Y}'_{\text{bound}} \log(q_t)$,
    with $\gamma=2$ and $\alpha_t>0.75$, focusing on hard examples and enhancing the gradient contribution from sparse positive samples.
    
    \item \textbf{Adaptive spatial penalty} prevents false positive predictions in non-boundary regions through a spatial constraint $\mathcal{L}_{p} = \sum_{j\in\Omega_{penalty}} \max(0, q_j - \tau_{penalty})$, where $\tau_{penalty} = \min\{\max(q_i)|i\in\Omega_{true}\}$ is dynamically derived from true boundary regions.
    
    \item \textbf{Classification loss} supervises the binary prediction head using BCEWithLogitsLoss, which applies sigmoid activation and binary cross-entropy directly to the raw outputs $\mathcal{L}_{c} = \text{BCEWithLogitsLoss}(\hat{\mathcal{Y}}_{\text{class}}, \mathcal{Y}_{\text{class}})$.
\end{itemize}

Where the focal loss $\mathcal{L}_f$ and adaptive spatial penalty loss $\mathcal{L}_{p}$ supervise the boundary prediction task, and the classification loss $\mathcal{L}_{c}$ handles the malware classification task.  
The final training of AV-DBNet is completed through multiple rounds of optimization aimed at minimizing these loss values.

\begin{figure*}[t]

  \centering
\captionsetup[subfigure]{skip=0pt}
  \captionsetup{skip=1pt}
  \begin{subfigure}[t]{0.480\textwidth} 

    \includegraphics[width=\linewidth]{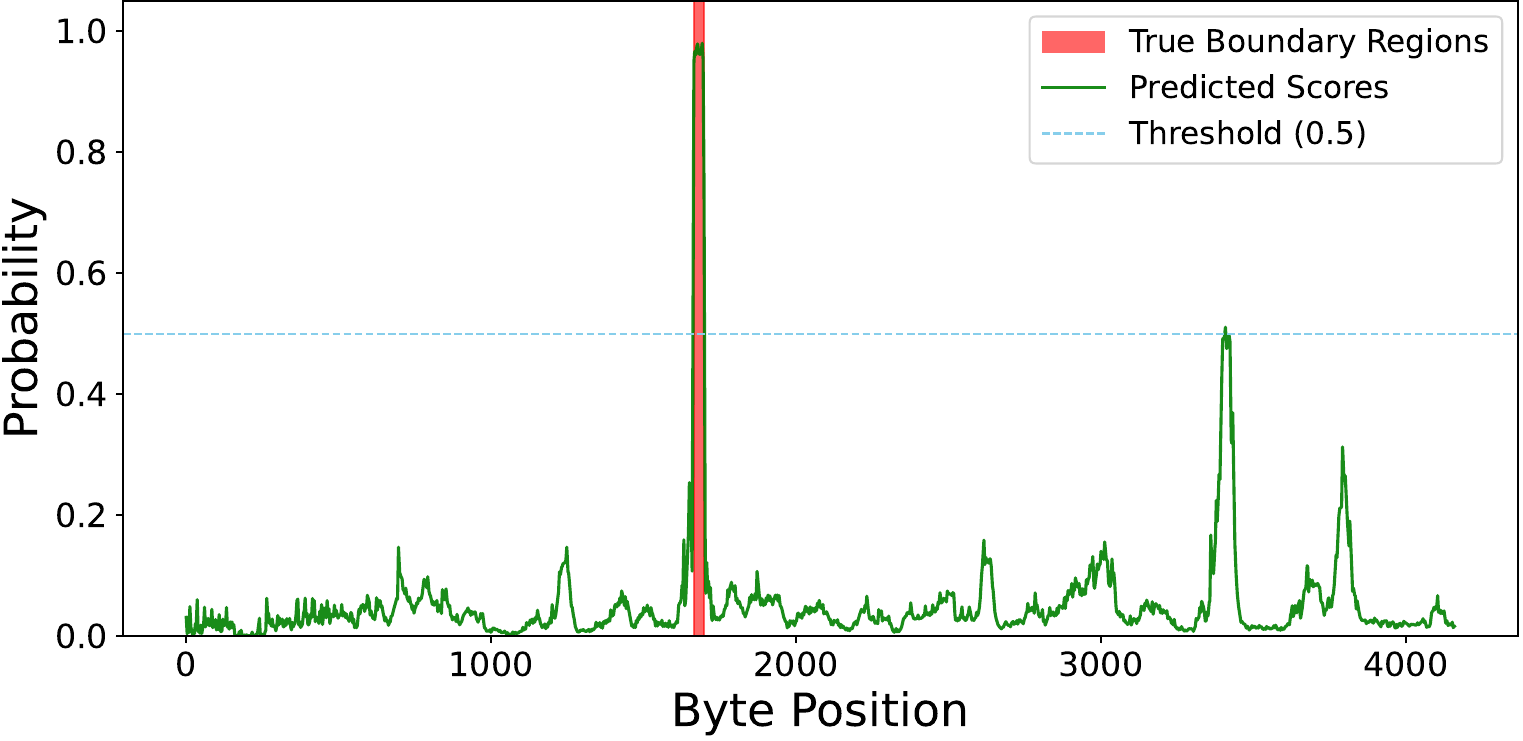}
    \caption{AV-DBNet (ClamAV Surrogate Version) Boundary Prediction on a Malware Sample\footnotemark with a Single Boundary Point. }
    \label{fig:one boundary}
       
\footnotetext{\fontsize{6.5pt}{7.5pt}\selectfont 1-SHA256: a468b9a6ce1c25a7e003e3f84a78b1ad69c9762a9d324ffdd9c16d0a33ded024}

  \end{subfigure}
  \hspace{0.01\textwidth} 
  \begin{subfigure}[t]{0.490\textwidth}
    \includegraphics[width=\linewidth]{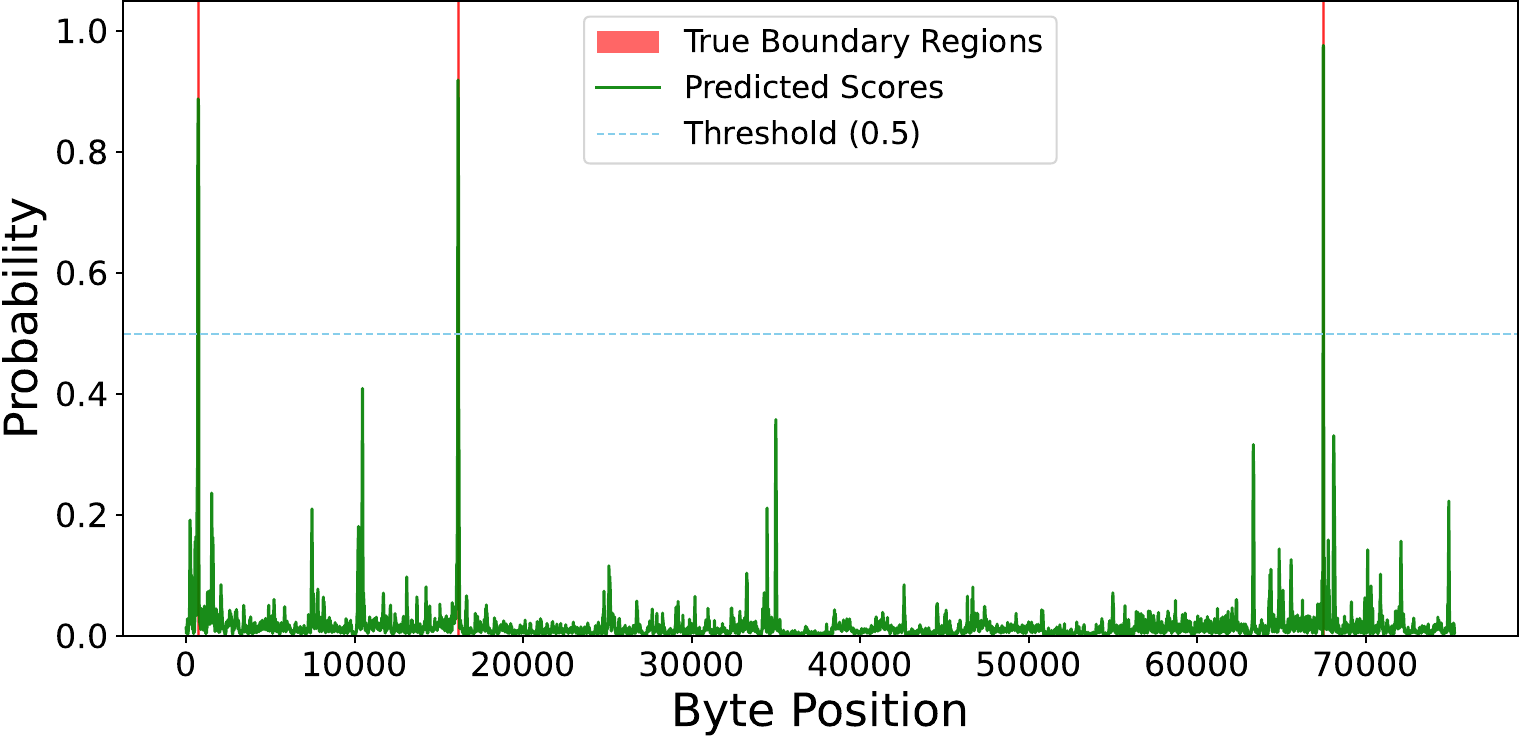}
    \caption{AV-DBNet (Avast Surrogate Version) Boundary Prediction on a Malware Sample\footnotemark with Multiple Boundary Points.}
    \label{fig:muti boundary}
\footnotetext{\fontsize{6.5pt}{7.5pt}\selectfont 2-SHA256: 4d0f327468402fba6b7ea4204f48bdc3b055b6c3b893e307506ffe83acaa699b}

  \end{subfigure}

  \caption{ Boundary prediction probability distributions produced by AV-DBNet. Probability peaks above $\tau_p=0.5$ that overlap with ground-truth regions indicate precise predictions.}

  \label{fig:combine predict}
\end{figure*}


\subsubsection{Model Outputs}

A trained AV-DBNet model produces two key outputs for a given executable: a binary classification score \(\hat{\mathcal{Y}}_{\text{class}}\) and a boundary probability distribution \(\hat{\mathcal{Y}}_{\text{bound}}\).

\textbf{Classification Output:}
The binary prediction $\hat{\mathcal{Y}}_{\text{class}}$ is obtained by thresholding the output score with $\tau_c$ (default $0.5$): $\text{Decision}=\mathbb{I}[\hat{\mathcal{Y}}_{\text{class}}>\tau_c]$. This output allows AV-DBNet to act as a high-agreement surrogate for the target AV. Since the classifier is jointly trained with byte-level boundary supervision, it remains more robust to injection-based perturbations that mislead conventional surrogates by focusing on boundary-critical regions. Quantitative results are provided in Section~\ref{Comparison with Existing Binary Detection Modeling Methods}.

\textbf{Boundary Region Inference from Probability Distributions:}
\label{Section:Boundary Region Inference from Probability Distributions}
Beyond binary classification, AV-DBNet outputs a boundary probability distribution $\hat{\mathcal{Y}}_{\text{bound}}$ that assigns each byte a likelihood of belonging to an AV decision boundary. This output enables explicit localization of decision-critical byte regions.
Predicted boundary regions are derived from $\hat{\mathcal{Y}}_{\text{bound}}$ through a two-step pipeline: 
\begin{itemize}
    \item Apply a threshold $\tau_p$ (default 0.5) to generate a binary mask $M_{\text{bin}}$, where $M_{\text{bin}}[i] = 1$ if $\hat{\mathcal{Y}}_{\text{bound}}[i] > \tau_p$.   
    \item  Contiguous ``1'' segments are merged to form the predicted boundary regions: $\hat{\text{BRs}} = \bigcup_i [\text{start}_i, \text{end}_i]$
\end{itemize}

By covering most of the ground-truth BPs, the predicted distribution $\hat{\mathcal{Y}}_{\text{bound}}$ enables effective inference of the target AV’s byte-level boundaries (see Section~\ref{Section:Boundary Prediction Capability}). The probability peaks above $\tau_p$ that overlap with ground-truth BRs can be visualized on a 2D coordinate plot (Figure~\ref{fig:combine predict}). This probability distribution therefore not only supports explicit localization of decision-critical byte regions, but also can provide precise boundary knowledge for downstream tasks such as boundary-guided evasion analysis, AV detection logic analysis, and cross-AV knowledge transfer assessment.

\section{Evaluation}

We evaluate AVHunter to assess its effectiveness in inferring AV decision boundaries, constructing high-fidelity surrogate models, and enabling robust evasion under realistic threat assumptions. Our evaluation is designed to measure both the accuracy and impact of AVHunter, as well as its robustness over time. To this end, we structure our experiments around the following research questions:

\begin{itemize}

    \item{\textbf{RQ1: Boundary Prediction Capability.}} How effectively can AVHunter predict BPs constituting the decision boundaries for malware samples across the 11 AVs?
    
    \item{\textbf{RQ2: Comparison with Binary-Feedback-Only Baselines.}} How does AVHunter compare with binary-feedback-only baselines in detection agreement and adversarial resilience?
    
    \item{\textbf{RQ3: Temporal Robustness.}} How stable are AVHunter’s surrogate models and boundary predictions over time as the target AVs evolve?

    \item{\textbf{RQ4: Ablation Analysis.}} How do predicted boundary signals contribute to AVHunter's surrogate agreement and adversarial detection performance across the 11 AVs?

\end{itemize}

\subsection{Evaluation Setup}

All experiments are conducted on BABD (Section~\ref{Section:Byte-Level AV Boundary Dataset}), which contains \texttt{241,070} malware and benign samples annotated with byte-level decision boundaries for 11 AV products. We split BABD into training and testing sets at a 5:1 ratio, stratified by malware type to ensure a fair evaluation across different sample categories. For each AV, its AV-specific training subset is used to train a dedicated AV-DBNet surrogate, and the corresponding testing subset is used for evaluation.

\subsubsection{Target Antivirus}
\label{Section:4.1.2}

To support large-scale automated evaluation, we select AV products that can run on Windows and provide reliable command-line detection interfaces consistent with their GUI behavior. In total, we evaluate 11 real-world AV products, including one open-source AV, ClamAV~\cite{ClamAV2026}, and 10 commercial AV products: Windows Defender~\cite{Windows2026Defender}, Kaspersky~\cite{Kaspersky2026}, ESET~\cite{ESET2026}, Avast~\cite{Avast2026}, AVG~\cite{AVG2026}, Vba32~\cite{Vba322026}, DrWeb~\cite{DrWeb2026}, IkarusT3~\cite{Ikarus2026}, FProtect~\cite{fprot_nlcv}, and McAfee~\cite{McAfee2026}. Among them, six have significant market share~\cite{Security2025Antivirus,Cybernews2025Antivirus}. For IkarusT3 and McAfee, whose desktop clients do not expose command-line interfaces, we use their corresponding offline command-line tools. All AV products are evaluated under default configurations and without query-rate limits during the subscription period.

\subsubsection{Baseline Methods}

We compare AVHunter with two types of baselines. For surrogate modeling, we adopt the two best-performing models from the state-of-the-art AV model-stealing work~\cite{rigaki2023stealing}, namely FFNN-TL and dualFFNN. Both are EMBER-feature-based binary-feedback-only surrogates and are fully trained under the same black-box threat model as AVHunter, where only binary AV feedback is available and no internal detection logic or byte-level boundaries are exposed.
For boundary-knowledge validation, we include GAMMA~\cite{demetrio2021functionality} and MalGuise~\cite{ling2024wolf} as coarse-grained modification baselines. Both rely on binary detection feedback and have reported limited evasion against real-world AVs without explicitly localizing byte-level AV boundaries. This comparison highlights the gap between binary-feedback-driven modification and AVHunter's fine-grained boundary-guided modification.

\subsubsection{Evaluation Metrics}
\label{Section:Evaluation Metrics}

We use the following metrics to evaluate AVHunter and the baselines. Let $N$ be the number of samples, $x_i$ the $i$-th sample, and $y_i^{\text{true}}$ its ground-truth label.

\begin{itemize}
\item \textbf{Detection Agreement (Agr)} measures consistency between the surrogate and target AV:
$ \text{Agr} = \frac{1}{N}\sum \mathbb{I}(f_{\text{sur}}(x_i) = f_{\text{AV}}(x_i)) $.

\item \textbf{Accuracy (Acc)} measures surrogate accuracy against ground-truth labels:
$ \text{Acc} = \frac{1}{N}\sum \mathbb{I}(f_{\text{sur}}(x_i) = y_i^{\text{true}}) $.

\item \textbf{Boundary Prediction Recall (BPR)} measures the fraction of ground-truth BPs covered by predicted BRs:
$ \text{BPR} = \frac{\text{TP BPs}}{\text{TP BPs} + \text{FN BPs}} $.

\item \textbf{Boundary Prediction Precision (BPP)} measures the fraction of predicted boundary regions that overlap with ground-truth BPs:
$ \text{BPP} = \frac{\text{TP}_{\text{pred}}}{\text{TP}_{\text{pred}} + \text{FP}_{\text{pred}}} $.

\item \textbf{Boundary Region Coverage Ratio (BRCR)} measures the byte-level compactness of predicted BRs:
$ \text{BRCR} = \frac{\text{Bytes covered by predicted BRs}}{\text{Total Bytes of Samples}} $.
\end{itemize}

\subsubsection{Experimental Setup}

All experiments are conducted on a Windows 11 system with an NVIDIA GeForce RTX 4080 SUPER GPU and an Intel Core i7-14700K CPU. For each target AV, we train one AV-DBNet surrogate for 50 epochs and select the checkpoint with the lowest validation loss. Unless otherwise specified, all hyperparameters follow the default settings in our released implementation.

\begin{table*}[t]
\renewcommand{\arraystretch}{1.2}

	\caption{{Boundary Prediction Performance of AVHunter Surrogate Models for 11 AV Products (Unit: \%).}}
	\label{tab:Boundary Prediction Results}
    \centering
    \resizebox{1\linewidth}{!}{\normalsize
\begin{tabular}{ccccccccccccc}
\hline
\textbf{Metrics} & \textbf{ClamAV} & \textbf{Defender} & \textbf{Kaspersky} & \textbf{ESET} & \textbf{Avast} & \textbf{AVG} & \textbf{DrWeb} & \textbf{Vba32} & \textbf{IkarusT3} & \textbf{FProtect} & \textbf{McAfee} & \textbf{Average} \\ \hline
\textbf{BPR}     & 85.38           & 75.23             & 81.04              & 87.05         & 88.18          & 90.66        & 83.23          & 89.21          & 82.10             & 83.67             & 90.05           & 85.07            \\
\textbf{BPP}     & 75.72           & 36.39             & 52.40              & 71.91         & 77.67          & 77.81        & 61.43          & 73.12          & 54.07             & 72.43             & 54.28           & 64.29            \\
\textbf{F1}      & 80.26           & 49.05             & 63.65              & 78.76         & 82.60          & 83.75        & 70.69          & 80.37          & 65.20             & 77.65             & 67.73           & 72.70            \\
\textbf{BRCR}    & 0.06            & 0.08              & 0.05               & 0.06          & 0.04           & 0.04         & 0.03           & 0.02           & 0.07              & 0.02              & 0.05            & 0.05             \\ \hline
\end{tabular}
}
\end{table*}

\begin{table*}[t]
\renewcommand{\arraystretch}{1.3}
    \caption{{Evasion rate, perturbation rate (PR), and functionality preservation rate (FPR) of generic malware modification methods and AVHunter boundary-guided BR encryption (Unit: \%.)}}
	\label{tab:Malware Evasion Attack Utility Results}
    \centering
    \resizebox{1\linewidth}{!}{\LARGE
\begin{tabular}{clcccccccccccclcc}
\hline
\multirow{2}{*}{\textbf{Modification Method}} &  & \multicolumn{12}{c}{\textbf{Evasion Rate}}                                                                                                                                                                                 &  & \multirow{2}{*}{\textbf{\begin{tabular}[c]{@{}c@{}}Average\\ PR\end{tabular}}} & \multirow{2}{*}{\textbf{\begin{tabular}[c]{@{}c@{}}Average\\ FPR\end{tabular}}} \\ \cline{3-14}
                                              &  & \textbf{ClamAV} & \textbf{Defender} & \textbf{Kaspersky} & \textbf{ESET}  & \textbf{Avast} & \textbf{AVG}   & \textbf{DrWeb} & \textbf{Vba32} & \textbf{IkarusT3} & \textbf{FProtect} & \textbf{McAfee} & \textbf{Average} &  &                                                                                &                                                                                 \\ \hline
\textbf{GAMMA}                                &  & 3.92            & 3.57              & 8.86               & 35.61          & 1.83           & 2.19           & 3.03           & 9.44           & 2.73              & 6.03              & 26.56           & 9.43             &  & 345.37                                                                         & \textbf{91.67 }                                                                       \\
\textbf{MalGuise}                             &  & 6.36            & 19.00             & 26.30              & 12.52          & 5.25           & 5.40           & 40.05          & 55.22          & 7.27              & 21.74             & 30.45           & 20.87            &  & 11.31                                                                          & 81.25                                                                           \\
\textbf{BR Encryption (Ours)}                 &  & \textbf{75.01}  & \textbf{82.15}    & \textbf{86.10}     & \textbf{85.14} & \textbf{79.93} & \textbf{80.08} & \textbf{92.47} & \textbf{90.84} & \textbf{75.02}    & \textbf{85.8}     & \textbf{89.05}  & \textbf{83.78}   &  & \textbf{3.22 }                                                                          & \textbf{91.67  }                                                                         \\ \hline
\end{tabular}
}
\end{table*}

\begin{table*}[t]
\renewcommand{\arraystretch}{1.2}
	\caption{{False Positive Induction by Injecting AVHunter-Predicted Boundary Regions into Benign Executables (Unit: \%).}}
	\label{tab:Benign Reversion Attack}
    \centering
    \resizebox{1\linewidth}{!}{\LARGE
\begin{tabular}{ccccccccccccc}
\hline
\textbf{Metric}              & \textbf{ClamAV} & \textbf{Defender} & \textbf{Kaspersky} & \textbf{ESET} & \textbf{Avast} & \textbf{AVG} & \textbf{DrWeb} & \textbf{Vba32} & \textbf{IkarusT3} & \textbf{FProtect} & \textbf{McAfee} & \textbf{Average} \\ \hline
\textbf{False Positive Rate} & 93.41           & 97.72             & 55.09              & -             & 99.95          & 99.91        & 99.94          & 51.42          & 50.81             & -                 & 83.71           & 81.33            \\
\textbf{Perturbation Rate}   & 0.07            & 0.07              & 0.14               & -             & 0.07           & 0.07         & 0.7            & 0.91           & 0.14              & -                 & 0.14            & 0.26             \\ \hline
\end{tabular}
}
\end{table*}

\subsection{RQ1: Boundary Prediction Capability}
\label{Section:Boundary Prediction Capability}

\subsubsection{Boundary Prediction}

We evaluate AVHunter’s capability to infer fine-grained, byte-level decision boundaries of AV products. Using the boundary probability distributions produced by AV-DBNet, BRs are derived through post-processing as defined in Section~\ref{Section:Boundary Region Inference from Probability Distributions}. The prediction performance is evaluated using the following metrics: Boundary Prediction Recall (BPR), Boundary Prediction Precision (BPP), F1 score, and Boundary Region Coverage Ratio (BRCR) (see Section~\ref{Section:Evaluation Metrics} for formal definitions), where F1 summarizes the balance between BPR and BPP.

As shown in Table~\ref{tab:Boundary Prediction Results}, AVHunter achieves consistently high boundary prediction performance across all 11 evaluated AV products, with an average BPR of 85.07\%. BPR exceeds 85\% for six AV products and remains above 75\% for all targets, indicating that AVHunter can reliably infer the majority of critical detection boundaries despite substantial variation in AV detection strategies. We consider BPR as the primary metric of interest, since it directly reflects the extent to which the true decision boundary is captured.

In contrast, the average BPP is 64.29\%. We speculate that the relatively lower precision arises because the large-scale training data enables the model to learn generalized patterns across samples; consequently, AV-DBNet sometimes predicts additional bytes that belong to genuine (yet previously unprobed) BRs in certain samples. The average F1 score is 72.70\%, which is mainly limited by the relatively lower BPP rather than insufficient boundary coverage. Although BPP is lower than BPR, we regard this as an acceptable trade-off. Importantly, the predicted BRs remain highly compact, with an average BRCR of only 0.05\%, indicating that the inferred boundary regions occupy only a very small fraction of the input bytes. This result further demonstrates the sparsity of AV decision-critical bytes, while suggesting that the additional predictions remain limited in scope and may still provide useful cues for identifying potential BRs in future analysis.

\begin{table*}[t]
    \renewcommand{\arraystretch}{1.2}
    \caption{{Detection agreement and accuracy on 11 AV products (Unit: \%).}}
    \label{tab:Detection Agreement Results of Common Samples}
    \centering
    \tiny
    \resizebox{0.7\linewidth}{!}{
        \begin{tabular}{ccccccccc}
            \hline
            \multirow{2}{*}{\textbf{No.}}
            & \multirow{2}{*}{\textbf{AV Name}}
            & \textbf{AV-Self}
            & \multicolumn{2}{c}{\textbf{FFNN-TL}}
            & \multicolumn{2}{c}{\textbf{dualFFNN}}
            & \multicolumn{2}{c}{\textbf{AVHunter}} \\
            \cline{3-9}
            & & \textbf{Acc}
            & \textbf{Agr} & \textbf{Acc}
            & \textbf{Agr} & \textbf{Acc}
            & \textbf{Agr} & \textbf{Acc} \\
            \hline
            1  & \textbf{ClamAV}    & 93.92 & 92.93 & 89.69 & 92.45 & 88.83 & \textbf{96.76} & \textbf{93.30} \\
            2  & \textbf{Defender}  & 96.67 & 95.30 & 94.43 & 95.02 & 93.75 & \textbf{97.29} & \textbf{96.33} \\
            3  & \textbf{Kaspersky} & 97.58 & 96.39 & 95.20 & 96.51 & 95.44 & \textbf{98.33} & \textbf{97.28} \\
            4  & \textbf{ESET}      & 93.82 & 94.66 & 90.25 & 94.88 & 90.08 & \textbf{98.17} & \textbf{93.69} \\
            5  & \textbf{Avast}     & 93.92 & 93.87 & 90.54 & 93.53 & 89.61 & \textbf{97.20} & \textbf{93.26} \\
            6  & \textbf{AVG}       & 93.66 & 93.90 & 90.28 & 93.66 & 89.97 & \textbf{97.20} & \textbf{93.00} \\
            7  & \textbf{DrWeb}     & 96.78 & 95.66 & 94.94 & 95.92 & 95.21 & \textbf{97.76} & \textbf{96.33} \\
            8  & \textbf{Vba32}     & 83.96 & 93.21 & 80.40 & 92.26 & 78.18 & \textbf{95.82} & \textbf{82.89} \\
            9  & \textbf{IkarusT3}  & 97.01 & 96.30 & 94.72 & 96.08 & 94.38 & \textbf{98.38} & \textbf{96.61} \\
            10 & \textbf{FProtect}  & 66.22 & 94.81 & 61.81 & 94.45 & 61.37 & \textbf{97.88} & \textbf{65.73} \\
            11 & \textbf{McAfee}    & 92.47 & 94.37 & 89.45 & 94.70 & 89.24 & \textbf{96.93} & \textbf{93.06} \\
             -  & \textbf{Average}   & 91.46 & 94.67 & 88.34 & 94.50 & 87.82 & \textbf{97.43} & \textbf{91.04} \\
            \hline
        \end{tabular}
    }
\end{table*}

\begin{figure*}[h]\centering

	\centering
	\includegraphics[width=\linewidth]{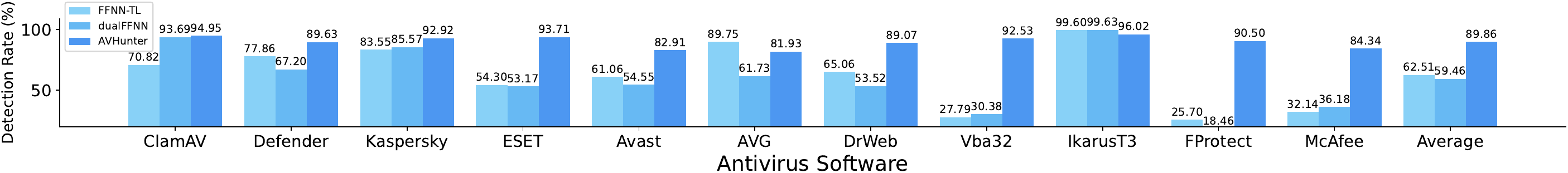}
    \caption{Detection rate of AVHunter and baseline methods on adversarially modified samples (Unit: \%).}
	\label{fig:Adversarial Detection Rates}
\end{figure*}

\subsubsection{Validating Boundary Predictions via AV Evasion}
\label{Section:Boundary-Guided Malware Modification}

To further validate whether the predicted BRs correspond to decision-critical content used by real-world AVs, we conduct a functionality-preserving boundary-coverage evaluation and measure the resulting evasion rate against the target AV products. For each AV target, the corresponding AV-specific AVHunter surrogate model, i.e., the AV-DBNet trained for that target AV, is used to provide detection-result feedback and boundary predictions. Specifically, \textbf{GAMMA}~\cite{demetrio2021functionality} and \textbf{MalGuise}~\cite{ling2024wolf} are used as two representative generic evasion methods that rely only on the binary-detection-result feedback of the corresponding AVHunter surrogate (whose detection performance is evaluated in Section~\ref{section:Detection Agreement}), without explicitly leveraging boundary localization; both have been shown in their original studies to evade real-world AV products. In contrast, \textbf{BR Encryption} is a boundary-aware method unique to AVHunter: it encrypts the predicted BRs and restores the original bytes at runtime through a decryption routine placed in a newly added section. Results are summarized in Table~\ref{tab:Malware Evasion Attack Utility Results}.

Across all 11 AV products, the two generic feedback-driven methods exhibit limited effectiveness on the testing set. Specifically, GAMMA and MalGuise achieve average evasion rates of 9.43\% and 20.87\%, respectively. In contrast, BR Encryption achieves a substantially higher average evasion rate of 83.78\% on the testing set, and consistently outperforms the other two methods across all evaluated AV products. These results show that AVHunter-guided boundary-aware modification is highly effective, thereby providing strong evidence that the boundary regions inferred by AVHunter closely match the true decision-critical content used by real-world AV products. Notably, for some AV targets (such as Windows Defender and Kaspersky), the evasion rate of BR Encryption is even higher than the corresponding BPR of the surrogate model. A possible explanation is that AVHunter sometimes predicts boundary regions that cover not only the probed ground-truth boundary points but also additional latent decision-critical bytes not captured during probing, which suggests a certain degree of generalization in its boundary inference.

In addition, BR Encryption requires much smaller modifications than the generic baselines. The average perturbation rates of GAMMA, MalGuise, and BR Encryption are 345.37\%, 11.31\%, and 3.22\%, respectively. This is because BR Encryption only encrypts the predicted BR bytes, enabling much more precise and localized modifications. Most of its perturbation overhead comes from the decryption routine inserted into the newly added section, rather than from large-scale rewriting of the original binary content. 

To assess functionality preservation, we further conduct an experiment on 48 collected known samples. Each sample is modified separately by the three methods, and the functionality preservation rate is calculated according to whether the modified sample remains functional. The results show that BR Encryption preserves functionality well, with a functionality preservation rate comparable to or higher than the generic baselines. This further indicates that AVHunter can localize compact yet highly influential boundary regions while preserving functionality.

\subsubsection{Validating Boundary Predictions via BR Injection}

We further validate the accuracy of AVHunter-predicted BRs by injecting them into section caves of benign executables and evaluating whether they can trigger false positives. As shown in Table~\ref{tab:Benign Reversion Attack}, AVHunter induces false positives in 9 out of 11 AV products, with an average false positive rate of 81.33\% and an average perturbation rate of only 0.26\%. For several targets, injecting a single predicted BR with a perturbation rate of only 0.07\% is sufficient to trigger high false positive rates between 93.41\% and 99.95\%. This suggests that the predicted BRs capture highly discriminative decision-critical byte patterns used in AV detection.

Some AV products require expanded context around a single BR to trigger detection, which explains their higher perturbation rates. For example, DrWeb reaches a 99.94\% false positive rate only after expanding one BR to 10 times its original context (PR = 0.7\%). No effective BRs are found for ESET and FProtect, likely because their detection requires additional conditions beyond a single injected BR. Overall, the results show that AVHunter-inferred BRs contain sufficient decision-critical information for most AV products.

\begin{tcolorbox}[
    colback=gray!20, 
    colframe=black!50,  
    boxrule=1pt,     
    arc=3mm,         
    boxsep=0mm,
    before skip=5pt,
    after skip=0pt,
    fonttitle=\bfseries 
]
\textbf{Answer to RQ1}:
AVHunter reliably localizes the BPs and BRs associated with AV decision-critical regions, achieving high boundary prediction recall while keeping the inferred BRs highly compact. The evasion and injection evaluations further show that the predicted BRs preserve decision-critical information used by real-world AV products, thereby supporting the accuracy and practical value of AVHunter's boundary inference.

\end{tcolorbox}

\begin{figure*}[t]
\centering
\begin{subfigure}{0.49\textwidth}
    \centering
    \includegraphics[width=\linewidth]{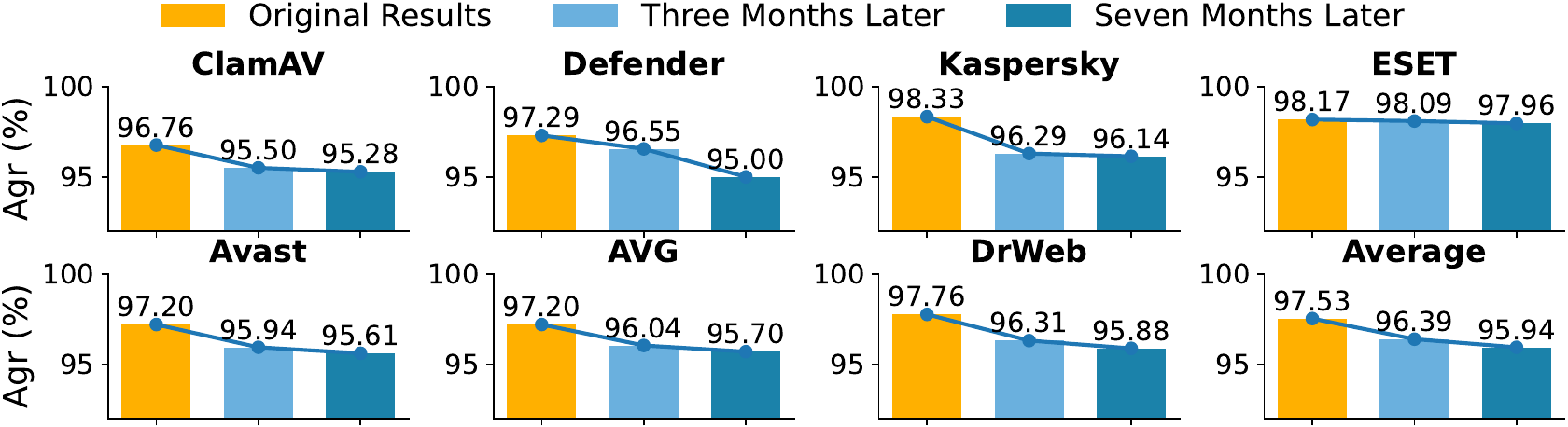}
    \caption{Surrogate Agreement Performance Degradation Results}
    \label{fig:Detection Agreement Degradation Results}
\end{subfigure}
\hfill
\begin{subfigure}{0.49\textwidth}
    \centering
    \includegraphics[width=\linewidth]{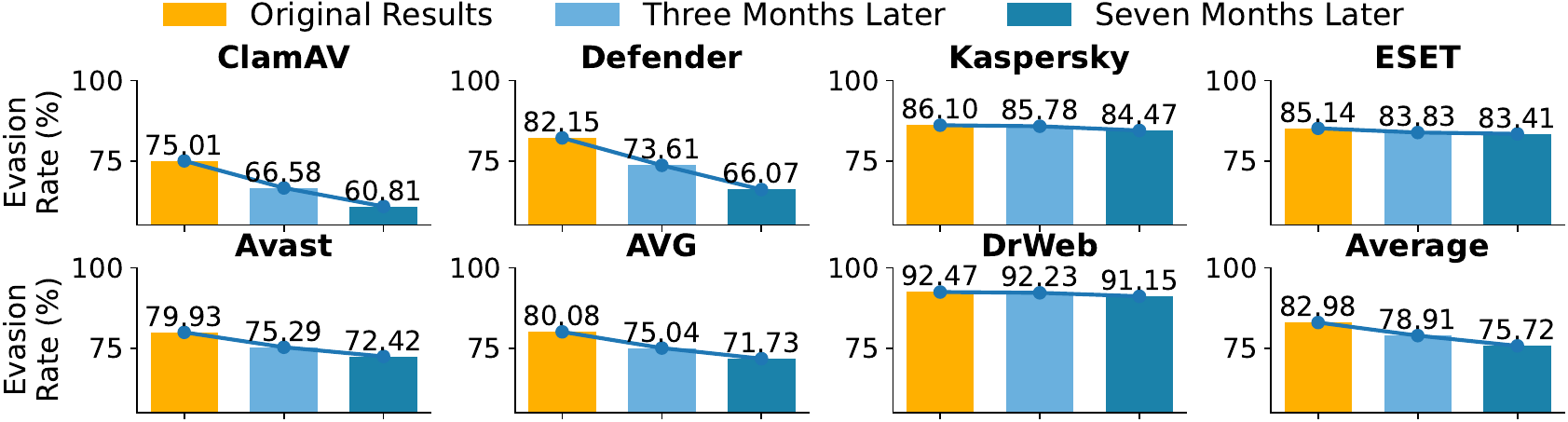}
    \caption{Malware Evasion Rate Degradation Results of BR Encryption}
    \label{fig:br_encryption_degradation_results}
\end{subfigure}

\vspace{0.1cm}

\begin{subfigure}{0.85\textwidth}
    \centering
    \includegraphics[width=\linewidth]{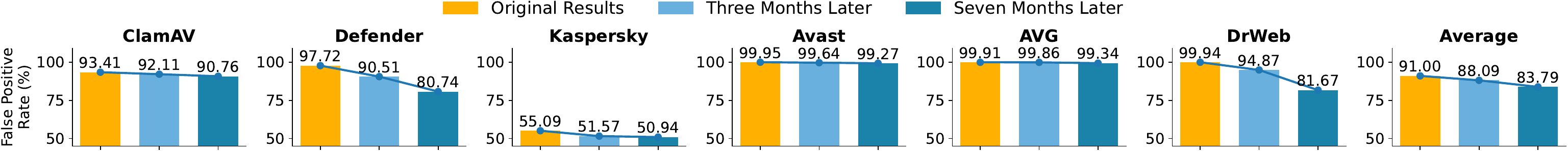}
    \caption{False Positive Rate Degradation Results of BR Injection}
    \label{fig:br_injection_degradation_results}
\end{subfigure}
\caption{Seven-Month Temporal Degradation of AVHunter on Continuously Updated AV Products Across Three Evaluation Stages: Initial, Three Months Later, and Seven Months Later.}
\label{fig:comprehensive_degradation_analysis}
\end{figure*}

\subsection{RQ2: Comparison with Binary-Feedback-Only Baselines}
\label{Comparison with Existing Binary Detection Modeling Methods}
\subsubsection{Detection Agreement}
\label{section:Detection Agreement}
We train AVHunter's surrogate model, AV-DBNet, and the baseline models, dualFFNN and FFNN-TL, using the 11 AV-specific training sets, and evaluate them on the corresponding testing sets (see Section~\ref{Section:Byte-Level AV Boundary Dataset}). The results are detailed in Table~\ref{tab:Detection Agreement Results of Common Samples}. AV-DBNet achieves the highest average detection agreement across the 11 target AVs, with an average agreement of 97.43\%, outperforming dualFFNN (94.50\%) and FFNN-TL (94.67\%). This indicates that explicit byte-level boundary supervision helps AV-DBNet capture AV-specific decision logic more effectively than binary-feedback-only baselines.

\subsubsection{Detection Resilience Against Adversarial Attacks} 
\label{Section:Detection Resilience Against Adversarial Attacks}

To evaluate adversarial resilience, we randomly select 2,000 test malware samples. For each surrogate model, we modify and evaluate only the samples that are originally classified correctly as malicious. Following prior malware evasion studies~\cite{demetrio2021functionality,ling2024wolf,he2024malwaretotal}, we instantiate a unified adversarial modification pipeline using common functionality-preserving transformations, including benign-byte section injection, control-flow-preserving rewriting, and PE metadata modification such as timestamps, version attributes, and section names. All surrogate models are evaluated under the same modification budget and configuration. The detection rate denotes the fraction of modified variants that remain classified as malicious.

As shown in Figure~\ref{fig:Adversarial Detection Rates}, AVHunter shows stronger overall robustness than the binary-feedback-only baselines under adversarial modifications. Across the 11 AV targets, AVHunter achieves the highest average detection rate of 89.86\%, compared with FFNN-TL (62.51\%) and dualFFNN (59.46\%). AVHunter also performs best on most AV targets and remains above 80\% in all cases. This resilience stems from AVHunter's boundary-aware modeling: by learning byte-level decision-critical regions beyond binary feedback, AV-DBNet can better preserve AV-specific malicious evidence under benign-byte injection, control-flow-preserving rewriting, and metadata modification, thereby reducing the impact of coarse-grained adversarial perturbations.

\begin{tcolorbox}[
    colback=gray!20, 
    colframe=black!50,  
    boxrule=1pt,     
    arc=3mm,         
    boxsep=0mm,
    before skip=5pt,
    after skip=0pt,
    fonttitle=\bfseries 
]
\textbf{Answer to RQ2.}
Compared with binary-feedback-only baselines, AVHunter achieves higher detection agreement and stronger adversarial resilience, demonstrating the benefit of explicit boundary-aware learning.
\end{tcolorbox}

\subsection{RQ3: Temporal Robustness}
\label{section:4.2.4}

\subsubsection{Surrogate Agreement Over Time}

To evaluate the temporal stability of AVHunter, we test its surrogate models on the BABD testing set against later versions of the same target AV products. This study covers seven AV products; the remaining four are excluded because they rely on offline tools or were discontinued, preventing consistent long-term evaluation.

As shown in Figure~\ref{fig:Detection Agreement Degradation Results}, AVHunter maintains high detection agreement over time. The average agreement across seven AV products decreases from 97.53\% at initialization to 96.39\% after three months and 95.94\% after seven months, corresponding to only 1.59\% average degradation. The degradation varies across AV products: Windows Defender shows the largest drop from 97.29\% to 95.00\%, while ESET remains highly stable with only 0.21\% decline. These results indicate that AVHunter remains robust to temporal drift in AV detection logic.

\subsubsection{Temporal Stability of BR Encryption and BR Injection Results}

We further evaluate whether AVHunter-inferred boundary regions remain effective after AV updates by re-testing samples generated by BR Encryption and BR Injection. As shown in Figures~\ref{fig:br_encryption_degradation_results} and~\ref{fig:br_injection_degradation_results}, both modification methods show only moderate degradation over seven months.

For BR Encryption, the largest drops occur on Windows Defender and ClamAV, with reductions of 16.08\% and 14.20\%, respectively, while the average evasion rate decreases by 7.26\%. For BR Injection, Windows Defender and Dr.Web show the most noticeable declines, at 16.98\% and 18.27\%, respectively, and the average false positive rate decreases from 91.00\% to 83.79\%, corresponding to 7.21\% degradation. Overall, these results show that many inferred BRs continue to preserve decision-critical information even as AV detection logic evolves.

\begin{tcolorbox}[
    colback=gray!20, 
    colframe=black!50,  
    boxrule=1pt,     
    arc=3mm,         
    boxsep=0mm,
    before skip=5pt,
    after skip=0pt,
    fonttitle=\bfseries 
]
\textbf{Answer to RQ3.}
AVHunter maintains high-fidelity surrogate behavior over time, and BR Encryption/Injection degrade only moderately across later AV versions, indicating that the inferred boundary regions remain largely stable as AV detection logic evolves.
\end{tcolorbox}

\begin{table*}[t]
\renewcommand{\arraystretch}{1.2}
\caption{Ablation study using all-zero, random, and predicted boundary masks. Metrics are surrogate agreement on unmodified samples and detection rate on adversarially modified samples (Unit: \%).}
\label{tab:Ablation Analysis Results of Variants 1-3}
\centering
\resizebox{0.95\linewidth}{!}{\LARGE
\begin{tabular}{cccccccccccccc}
\hline
\textbf{Ablation Settings} & \textbf{Metrics} & \textbf{ClamAV} & \textbf{Defender} & \textbf{Kaspersky} & \textbf{ESET} & \textbf{Avast} & \textbf{AVG} & \textbf{DrWeb} & \textbf{Vba32} & \textbf{IkarusT3} & \textbf{FProtect} & \textbf{McAfee} & \textbf{Average} \\ \hline
\multirow{2}{*}{\textbf{No Boundary Signal}}
& \textbf{Agreement} & 89.79 & 93.01 & 92.91 & 92.69 & 92.03 & 86.18 & 87.37 & 74.83 & 96.93 & 89.39 & 92.58 & 89.79 \\
& \textbf{Detection Rate} & 73.96 & 83.86 & 66.33 & 61.60 & 59.49 & 51.25 & 37.79 & 8.68 & 88.48 & 13.58 & 70.40 & 55.95 \\ \hline
\multirow{2}{*}{\textbf{Random Boundary Signal}}
& \textbf{Agreement} & 91.91 & 94.03 & 94.49 & 94.67 & 92.93 & 88.36 & 89.69 & 80.23 & 96.93 & 90.03 & 91.35 & 91.33 \\
& \textbf{Detection Rate} & 76.46 & 84.53 & 70.12 & 67.17 & 62.60 & 53.03 & 42.18 & 14.41 & 88.37 & 13.75 & 70.73 & 58.49 \\ \hline
\multirow{2}{*}{\textbf{Full Model}}
& \textbf{Agreement} & 96.76 & 97.29 & 98.33 & 98.17 & 97.20 & 97.20 & 97.76 & 95.82 & 98.38 & 97.88 & 96.93 & 97.43 \\
& \textbf{Detection Rate} & 94.95 & 89.63 & 92.92 & 93.71 & 82.91 & 81.93 & 89.07 & 92.53 & 96.02 & 90.50 & 84.34 & 89.86 \\ \hline
\end{tabular}
}
\par\smallskip

\end{table*}
\subsection{RQ4: Ablation Analysis}
\label{tab:Ablation Analysis Results}

To assess the contribution of predicted boundary information to AVHunter, we conduct an inference-time ablation study. We replace the predicted boundary mask while retaining the jointly trained parameters and classification architecture.

\subsubsection{Ablation Design}

We consider three settings:
\textbf{(1) No Boundary Signal} replaces the predicted boundary mask with an all-zero mask, producing zero-filled regional inputs;
\textbf{(2) Random Boundary Signal} replaces it with a binary mask sampled independently at each valid byte position from a Bernoulli distribution with probability $0.5$, leaving padding positions unselected;
\textbf{(3) Full Model} uses the predicted boundary mask to extract BR-based regional inputs alongside global raw-byte features.

Random regional inputs contain selected original file bytes rather than random byte values. Extraction follows the model's existing region-count and region-length limits, without matching random regions to predicted regions in size or count. We use random seed 2026 and a classification threshold of $0.5$. No model parameters are retrained.

Following the evaluation protocol of RQ2, we report surrogate agreement on unmodified samples and detection rate on adversarially modified samples. Results are averaged equally across the 11 AV models. This experiment examines the use of boundary information at inference time rather than the independent benefit of boundary supervision during training.

\subsubsection{Impact of Boundary Knowledge}

As shown in Table~\ref{tab:Ablation Analysis Results of Variants 1-3}, the Full Model achieves 97.43\% average agreement and 89.86\% adversarial detection. In comparison, No Boundary Signal achieves 89.79\% agreement and 55.95\% adversarial detection, while Random Boundary Signal achieves 91.33\% and 58.49\%, respectively. The full-model results are higher than those of both variants across all 11 AV models.

The numerical differences are particularly pronounced for adversarial detection. This pattern suggests that predicted boundary signals provide useful regional information complementary to global raw-byte features, helping AVHunter recognize malicious evidence in adversarially modified samples.

Although random regional inputs perform slightly better than zero-filled inputs on average, they do not match the performance obtained with predicted boundary signals. This suggests that relevant regional information, rather than merely additional regional bytes, contributes to effective classification.

\begin{tcolorbox}[
    colback=gray!20,
    colframe=black!50,
    boxrule=1pt,
    arc=3mm,
    boxsep=0mm,
    before skip=5pt,
    after skip=0pt,
    fonttitle=\bfseries
]
\textbf{Answer to RQ4}:
The results suggest that AVHunter's predicted boundary signals provide useful information for classification, particularly for detecting adversarially modified samples. This information may help the model maintain detection performance under adversarial modifications, thereby improving its robustness.
\end{tcolorbox}

\section{Related Work}
\label{Section:Related Work}

\textbf{Model Stealing and Surrogate Modeling.}
Model stealing has been widely studied in computer vision and natural language processing, where attackers query black-box models and train high-fidelity surrogates to imitate their decision behavior~\cite{chen2023d,sha2023can,zhao2024fully,zhuang2025stealix,luan2025dynamic,yuan2024data,ICLR2025_cce0e917,karmakar2023marich,liu2025model,naseh2023stealing}. Prior work further shows that such attacks can extend to malware detection and real-world AV products, achieving high detection agreement through black-box queries and semi-supervised learning~\cite{rigaki2023stealing}. However, these methods mainly approximate global decision behavior from coarse binary feedback and do not explicitly recover byte-level decision-critical regions, which are essential for understanding the discrete detection logic of real-world AV products.

\textbf{Black-Box Malware Modification.}
Another line of work studies malware modification under black-box detector feedback. Reinforcement-learning-based methods~\cite{anderson2018learning,zhao2021structural,molloy2022h4rm0ny,zhong2022reinforcement,song2022mab,zhang2023semantics,kozak2024creating,jha2023codeattack,zapzalka2024semantics,he2024malwaretotal} and deep-learning-based methods~\cite{hu2022generating,chen2020android,yefet2020adversarial,pierazzi2020intriguing,lucas2021malware,li2023black,lucas2023adversarial} attempt to infer detector behavior and generate functionality-preserving adversarial samples. However, their inferred boundaries are typically implicit and coarse-grained, because their primary goal is evasion rather than explicit localization of decision-critical bytes. This limitation is particularly evident against real-world AV products, where prior black-box approaches remain much less effective~\cite{rigaki2023stealing,demetrio2021functionality,ling2024wolf}.

AVHunter targets byte-level boundary inference for real-world AV products under black-box constraints. By probing boundary points, constructing AV-specific boundary labels, and training a surrogate model that jointly learns classification behavior and boundary regions, AVHunter infers static, pattern-oriented AV boundary knowledge. This suggests that model knowledge leakage is not limited to ML-to-ML imitation: discrete AV detection knowledge can also be mapped into learned feature spaces, potentially supporting unauthorized surrogate services, boundary-guided transfer attacks, and future integration with ML-based malware detection knowledge.

\section{Discussion}

\subsection{Limitation and Future Work}

Despite its strong empirical performance, AVHunter has several limitations. First, AVHunter focuses on static AV detection boundaries, so samples may still be detected by dynamic execution, sandbox analysis, or cloud-based analysis; extending AVHunter to dynamic boundaries remains important. Second, AVHunter depends on the scale and diversity of BABD, and focusing on samples with fewer than 20 BPs improves feasibility but may bias evaluation toward simpler detection logic. Future work should expand dataset coverage and develop probing strategies for complex or rare AV decision rules. Third, boundary probing introduces query overhead: AVHunter requires 19.20 queries per BP and 66.53 queries per malware sample on average, although this remains practical because most evaluated AV products impose no strict query limits and AV boundaries show relative temporal stability. Finally, AVHunter shows that AV boundary knowledge can be mapped into learned feature spaces, motivating future work on integrating AV detection knowledge with ML-based malware detection knowledge and studying the risks of such transfer.

\subsection{Mitigations}
\label{Section:Mitigations}

To mitigate AVHunter-style attacks, AV vendors may consider the following countermeasures: (1) Pre-scan Integrity Verification. Introducing integrity or anomaly checks prior to full scanning can detect suspicious probing behavior and return non-deterministic or misleading responses.
(2) Boundary Obfuscation. Employing multiple independent pattern sets and randomly selecting among them during detection can obscure decision boundaries and reduce the precision of boundary probing.
(3) Accelerated Signature Updates. Increasing the update frequency of detection signatures can exploit AVHunter’s temporal sensitivity and invalidate learned boundary information more rapidly. Some of these defenses may introduce trade-offs between detection accuracy and operational efficiency, warranting careful evaluation by AV vendors.

\subsection{Ethical Consideration}

This work studies the robustness of real-world AV systems through black-box boundary inference and surrogate modeling. Although AVHunter can guide malware modifications in controlled settings, our study is strictly limited to defensive security analysis and system measurement.
All experiments are conducted in isolated environments using offline or sandboxed AV engines. We do not release, deploy, or test malware in the wild, nor do we publicly distribute surrogate models or boundary annotations in a misuse-enabling form. The threat model assumes only realistic black-box access already considered by AV vendors.
Our goal is to expose structural weaknesses in current AV detection logic, particularly its reliance on sparse and stable patterns, so that defenders can better assess risks and improve resilience. We believe that measuring and responsibly disclosing these limitations provides defensive value and aligns with responsible security research practices.

\section{Conclusion}

This paper presents AVHunter, the first framework for inferring fine-grained byte-level decision-critical regions of real-world antivirus products under a black-box threat model. By constructing BABD, the first large-scale byte-level AV boundary dataset, we show that many AV detections are associated with a small number of compact decision-critical byte regions. Leveraging this insight, AVHunter not only reproduces binary AV decisions with high agreement, but also accurately localizes the byte regions underlying these decisions across 11 real-world AV products. We further demonstrate that the predicted regions preserve genuine AV decision knowledge through boundary-guided malware evasion, false-positive induction, and temporal evaluation. AVHunter outperforms binary-label surrogate baselines in AV agreement and adversarial resilience, while its boundary-guided modifications achieve stronger evasion than generic adversarial baselines. These findings suggest that AV knowledge leakage extends beyond malware/benign decisions: fine-grained boundary knowledge can also be recovered from black-box interactions. We hope this work motivates future research on boundary-aware malware analysis, robust AV design, and defenses against decision-boundary leakage.



\bibliographystyle{IEEEtran}
\bibliography{refs}

@article{rigaki2023stealing,
  title={Stealing and evading malware classifiers and antivirus at low false positive conditions},
  author={Rigaki, Maria and Garcia, Sebastian},
  journal={Computers \& Security},
  volume={129},
  pages={103192},
  year={2023},
  publisher={Elsevier}
}

@article{cavallaro2023machine,
  title={Are machine learning models for malware detection ready for prime time?},
  author={Cavallaro, Lorenzo and Kinder, Johannes and Pendlebury, Feargus and Pierazzi, Fabio},
  journal={IEEE Security \& Privacy},
  volume={21},
  number={2},
  pages={53--56},
  year={2023},
  publisher={IEEE}
}

@inproceedings{yang2021bodmas,
  title={BODMAS: An open dataset for learning based temporal analysis of PE malware},
  author={Yang, Limin and Ciptadi, Arridhana and Laziuk, Ihar and Ahmadzadeh, Ali and Wang, Gang},
  booktitle={2021 IEEE Security and Privacy Workshops (SPW)},
  pages={78--84},
  year={2021},
  organization={IEEE}
}

@article{harang2020sorel,
  title={SOREL-20M: A large scale benchmark dataset for malicious PE detection},
  author={Harang, Richard and Rudd, Ethan M},
  journal={arXiv preprint arXiv:2012.07634},
  year={2020}
}

@misc{Windows2026Defender,
  title = {{Windows Defender}},
  author = {{Microsoft}},
  year = {2026},
  note = {Accessed: 2026-06-29},
  howpublished = {\url{https://www.microsoft.com/en-us/windows/comprehensive-security}}
}

@misc{Kaspersky2026,
  title = {{Kaspersky}},
  author = {{Kaspersky Lab}},
  year = {2026},
  note = {Accessed: 2026-06-29},
  howpublished = {\url{https://www.kaspersky.com/}}
}

@misc{ESET2026,
  title = {{ESET Internet Security}},
  author = {{ESET}},
  year = {2026},
  note = {Accessed: 2026-06-29},
  howpublished = {\url{https://www.eset.com/}}
}

@misc{Avast2026,
  title = {{Avast Antivirus}},
  author = {{Avast Software}},
  year = {2026},
  note = {Accessed: 2026-06-29},
  howpublished = {\url{https://www.avast.com/}}
}

@misc{AVG2026,
  title = {{AVG Antivirus}},
  author = {{AVG Technologies}},
  year = {2026},
  note = {Accessed: 2026-06-29},
  howpublished = {\url{https://www.avg.com/}}
}

@misc{Vba322026,
  title = {{Vba32 Antivirus}},
  author = {{Vba32}},
  year = {2026},
  note = {Accessed: 2026-06-29},
  howpublished = {\url{https://www.anti-virus.by/}}
}

@misc{DrWeb2026,
  title = {{Dr.Web}},
  author = {{Doctor Web}},
  year = {2026},
  note = {Accessed: 2026-06-29},
  howpublished = {\url{https://www.drweb.cn/}}
}

@misc{Ikarus2026,
  title = {{Ikarus Antivirus}},
  author = {{Ikarus Security Software}},
  year = {2026},
  note = {Accessed: 2026-06-29},
  howpublished = {\url{https://www.ikarussecurity.com/}}
}

@misc{fprot_nlcv,
  title = {{F-Prot Antivirus for Windows}},
  author = {{National Laboratory of Computer Virology, Bulgarian Academy of Sciences}},
  howpublished = {\url{https://old.nlcv.bas.bg/?p=35&l=2}},
  note = {Product/reseller page for F-Prot Antivirus. Accessed: 2026-06-29},
  year = {2026},
}

@misc{McAfee2026,
  title = {{McAfee Antivirus}},
  author = {{McAfee LLC}},
  year = {2026},
  note = {Accessed: 2026-06-29},
  howpublished = {\url{https://www.mcafee.com}}
}

@misc{ClamAV2026,
  title = {{ClamAV}},
  author = {{Cisco}},
  year = {2026},
  note = {Accessed: 2026-06-29},
  howpublished = {\url{https://www.clamav.net}}
}

@misc{virustotal,
  title = {{VirusTotal}},
  author = {{VirusTotal}},
  year = {2026},
  note = {Accessed: 2026-06-29},
  howpublished = {\url{https://www.virustotal.com/}}
}

@misc{virusshare,
  title = {{VirusShare}},
  author = {{VirusShare}},
  year = {2002--2026},
  note = {Accessed: 2026-06-29},
  howpublished = {\url{https://virusshare.com/}}
}

@article{demetrio2021functionality,
  title={Functionality-preserving black-box optimization of adversarial windows malware},
  author={Demetrio, Luca and Biggio, Battista and Lagorio, Giovanni and Roli, Fabio and Armando, Alessandro},
  journal={IEEE Transactions on Information Forensics and Security},
  volume={16},
  pages={3469--3478},
  year={2021},
  publisher={IEEE}
}

@inproceedings{ling2024wolf,
  title={A wolf in sheep's clothing: practical black-box adversarial attacks for evading learning-based windows malware detection in the wild},
  author={Ling, Xiang and Wu, Zhiyu and Wang, Bin and Deng, Wei and Wu, Jingzheng and Ji, Shouling and Luo, Tianyue and Wu, Yanjun},
  booktitle={33rd USENIX Security Symposium (USENIX Security 24)},
  pages={7393--7410},
  year={2024}
}

@inproceedings{wressnegger2017automatically,
  title={Automatically inferring malware signatures for anti-virus assisted attacks},
  author={Wressnegger, Christian and Freeman, Kevin and Yamaguchi, Fabian and Rieck, Konrad},
  booktitle={Proceedings of the 2017 ACM on Asia conference on computer and communications security},
  pages={587--598},
  year={2017}
}

@article{anderson2018learning,
  title={Learning to evade static PE machine learning malware models via reinforcement learning},
  author={Anderson, Hyrum S and Kharkar, Anant and Filar, Bobby and Evans, David and Roth, Phil},
  journal={arXiv preprint arXiv:1801.08917},
  year={2018}
}

@inproceedings{molloy2022h4rm0ny,
  title={H4rm0ny: A competitive zero-sum two-player markov game for multi-agent learning on evasive malware generation and detection},
  author={Molloy, Christopher and Ding, Steven HH and Fung, Benjamin CM and Charland, Philippe},
  booktitle={2022 IEEE International Conference on Cyber Security and Resilience (CSR)},
  pages={22--29},
  year={2022},
  organization={IEEE}
}

@inproceedings{he2024malwaretotal,
  title={MalwareTotal: Multi-faceted and sequence-aware bypass tactics against static malware detection},
  author={He, Shuai and Fu, Cai and Hu, Hong and Chen, Jiahe and Lv, Jianqiang and Jiang, Shuai},
  booktitle={Proceedings of the IEEE/ACM 46th International Conference on Software Engineering},
  pages={1--12},
  year={2024}
}

@inproceedings{zhao2021structural,
  title={Structural attack against graph based android malware detection},
  author={Zhao, Kaifa and Zhou, Hao and Zhu, Yulin and Zhan, Xian and Zhou, Kai and Li, Jianfeng and Yu, Le and Yuan, Wei and Luo, Xiapu},
  booktitle={Proceedings of the 2021 ACM SIGSAC conference on computer and communications security},
  pages={3218--3235},
  year={2021}
}

@article{zhong2022reinforcement,
  title={Reinforcement learning based adversarial malware example generation against black-box detectors},
  author={Zhong, Fangtian and Hu, Pengfei and Zhang, Guoming and Li, Hong and Cheng, Xiuzhen},
  journal={Computers \& Security},
  volume={121},
  pages={102869},
  year={2022},
  publisher={Elsevier}
}

@inproceedings{song2022mab,
  title={MAB-Malware: A reinforcement learning framework for blackbox generation of adversarial malware},
  author={Song, Wei and Li, Xuezixiang and Afroz, Sadia and Garg, Deepali and Kuznetsov, Dmitry and Yin, Heng},
  booktitle={Proceedings of the 2022 ACM on Asia conference on computer and communications security},
  pages={990--1003},
  year={2022}
}

@ARTICLE{zhang2023semantics,
  author={Zhang, Lan and Liu, Peng and Choi, Yoon-Ho and Chen, Ping},
  journal={IEEE Transactions on Dependable and Secure Computing}, 
  title={Semantics-Preserving Reinforcement Learning Attack Against Graph Neural Networks for Malware Detection}, 
  year={2023},
  volume={20},
  number={2},
  pages={1390-1402},
  doi={10.1109/TDSC.2022.3153844}
  
  }

@article{kozak2024creating,
  title={Creating valid adversarial examples of malware},
  author={Koz{\'a}k, Matou{\v{s}} and Jure{\v{c}}ek, Martin and Stamp, Mark and Troia, Fabio Di},
  journal={Journal of Computer Virology and Hacking Techniques},
  volume={20},
  number={4},
  pages={607--621},
  year={2024},
  publisher={Springer}
}

@inproceedings{jha2023codeattack,
  title={Codeattack: Code-based adversarial attacks for pre-trained programming language models},
  author={Jha, Akshita and Reddy, Chandan K},
  booktitle={Proceedings of the AAAI Conference on Artificial Intelligence},
  volume={37},
  number={12},
  pages={14892--14900},
  year={2023}
}

@InProceedings{hu2022generating,
author="Hu, Weiwei and Tan, Ying",
editor="Tan, Ying and Shi, Yuhui",
title="Generating Adversarial Malware Examples for Black-Box Attacks Based on GAN",
booktitle="Data Mining and Big Data",
year="2022",
publisher="Springer Nature Singapore",
address="Singapore",
pages="409--423",
doi={10.1007/978-981-19-8991-9_29}
}

@ARTICLE{chen2020android,
  author={Chen, Xiao and Li, Chaoran and Wang, Derui and Wen, Sheng and Zhang, Jun and Nepal, Surya and Xiang, Yang and Ren, Kui},
  journal={IEEE Transactions on Information Forensics and Security}, 
  title={Android HIV: A Study of Repackaging Malware for Evading Machine-Learning Detection}, 
  year={2020},
  volume={15},
  number={},
  pages={987-1001},
  doi={10.1109/TIFS.2019.2932228}
  }

@article{yefet2020adversarial,
  title={Adversarial examples for models of code},
  author={Yefet, Noam and Alon, Uri and Yahav, Eran},
  journal={Proceedings of the ACM on Programming Languages},
  volume={4},
  number={OOPSLA},
  pages={1--30},
  year={2020},
  publisher={ACM New York, NY, USA}
}

@inproceedings{pierazzi2020intriguing,
  title={Intriguing properties of adversarial ml attacks in the problem space},
  author={Pierazzi, Fabio and Pendlebury, Feargus and Cortellazzi, Jacopo and Cavallaro, Lorenzo},
  booktitle={2020 IEEE symposium on security and privacy (SP)},
  pages={1332--1349},
  year={2020},
  organization={IEEE}
}

@inproceedings{lucas2021malware,
  title={Malware makeover: Breaking ml-based static analysis by modifying executable bytes},
  author={Lucas, Keane and Sharif, Mahmood and Bauer, Lujo and Reiter, Michael K and Shintre, Saurabh},
  booktitle={Proceedings of the 2021 ACM Asia Conference on Computer and Communications Security},
  pages={744--758},
  year={2021}
}

@inproceedings{li2023black,
  title={Black-box Adversarial Example Attack towards $\{$FCG$\}$ Based Android Malware Detection under Incomplete Feature Information},
  author={Li, Heng and Cheng, Zhang and Wu, Bang and Yuan, Liheng and Gao, Cuiying and Yuan, Wei and Luo, Xiapu},
  booktitle={32nd USENIX Security Symposium (USENIX Security 23)},
  pages={1181--1198},
  year={2023}
}

@inproceedings{lucas2023adversarial,
  title={Adversarial training for $\{$Raw-Binary$\}$ malware classifiers},
  author={Lucas, Keane and Pai, Samruddhi and Lin, Weiran and Bauer, Lujo and Reiter, Michael K and Sharif, Mahmood},
  booktitle={32nd USENIX Security Symposium (USENIX Security 23)},
  pages={1163--1180},
  year={2023}
}

@inproceedings{chen2023d,
  title={D-dae: Defense-penetrating model extraction attacks},
  author={Chen, Yanjiao and Guan, Rui and Gong, Xueluan and Dong, Jianshuo and Xue, Meng},
  booktitle={2023 IEEE Symposium on Security and Privacy (SP)},
  pages={382--399},
  year={2023},
  organization={IEEE}
}

@inproceedings{sha2023can,
  title={Can't steal? cont-steal! contrastive stealing attacks against image encoders},
  author={Sha, Zeyang and He, Xinlei and Yu, Ning and Backes, Michael and Zhang, Yang},
  booktitle={Proceedings of the IEEE/CVF Conference on Computer Vision and Pattern Recognition},
  pages={16373--16383},
  year={2023}
}

@inproceedings{zhao2024fully,
  title={Fully exploiting every real sample: Superpixel sample gradient model stealing},
  author={Zhao, Yunlong and Deng, Xiaoheng and Liu, Yijing and Pei, Xinjun and Xia, Jiazhi and Chen, Wei},
  booktitle={Proceedings of the IEEE/CVF Conference on Computer Vision and Pattern Recognition},
  pages={24316--24325},
  year={2024}
}

@InProceedings{zhuang2025stealix,
  title = 	 {Stealix: Model Stealing via Prompt Evolution},
  author =       {Zhuang, Zhixiong and Wang, Hui-Po and Nicolae, Maria-Irina and Fritz, Mario},
  booktitle = 	 {Proceedings of the 42nd International Conference on Machine Learning},
  pages = 	 {80551--80568},
  year = 	 {2025},
  editor = 	 {Singh, Aarti and Fazel, Maryam and Hsu, Daniel and Lacoste-Julien, Simon and Berkenkamp, Felix and Maharaj, Tegan and Wagstaff, Kiri and Zhu, Jerry},
  volume = 	 {267},
  series = 	 {Proceedings of Machine Learning Research},
  month = 	 {13--19 Jul},
  publisher =    {PMLR},


}

@inproceedings{
luan2025dynamic,
title={Dynamic Neural Fortresses: An Adaptive Shield for Model Extraction Defense},
author={Siyu Luan and Zhenyi Wang and Li Shen and Zonghua Gu and Chao Wu and Dacheng Tao},
booktitle={The Thirteenth International Conference on Learning Representations},
year={2025},
}

@inproceedings{yuan2024data,
  title={Data-free hard-label robustness stealing attack},
  author={Yuan, Xiaojian and Chen, Kejiang and Huang, Wen and Zhang, Jie and Zhang, Weiming and Yu, Nenghai},
  booktitle={Proceedings of the AAAI Conference on Artificial Intelligence},
  volume={38},
  number={7},
  pages={6853--6861},
  year={2024}
}

@inproceedings{ICLR2025_cce0e917,
 author = {Nasr, Milad and Rando, Javier and Carlini, Nicholas and Hayase, Jonathan and Jagielski, Matthew and Cooper, A. Feder and Ippolito, Daphne and Choquette-Choo, Christopher and Tramer, Florian and Lee, Katherine},
 booktitle = {International Conference on Learning Representations},
 editor = {Y. Yue and A. Garg and N. Peng and F. Sha and R. Yu},
 pages = {82363--82435},
 title = {Scalable Extraction of Training Data from Aligned, Production Language Models},
 volume = {2025},
 year = {2025}
}

@article{karmakar2023marich,
  title={Marich: A query-efficient distributionally equivalent model extraction attack},
  author={Karmakar, Pratik and Basu, Debabrota},
  journal={Advances in Neural Information Processing Systems},
  volume={36},
  pages={72412--72445},
  year={2023}
}

@inproceedings{liu2025model,
  title={Model stealing for any low-rank language model},
  author={Liu, Allen and Moitra, Ankur},
  booktitle={Proceedings of the 57th Annual ACM Symposium on Theory of Computing},
  pages={1755--1761},
  year={2025}
}

@inproceedings{naseh2023stealing,
  title={Stealing the decoding algorithms of language models},
  author={Naseh, Ali and Krishna, Kalpesh and Iyyer, Mohit and Houmansadr, Amir},
  booktitle={Proceedings of the 2023 ACM SIGSAC Conference on Computer and Communications Security},
  pages={1835--1849},
  year={2023}
}

@ARTICLE{zapzalka2024semantics,
  author={Zapzalka, Dylan and Salem, Saeed and Mohaisen, David},
  journal={IEEE Transactions on Dependable and Secure Computing}, 
  title={Semantics-Preserving Node Injection Attacks Against GNN-Based ACFG Malware Classifiers}, 
  year={2025},
  volume={22},
  number={1},
  pages={549-560},
  doi={10.1109/TDSC.2024.3409410}
  }

@inproceedings{lin2017focal,
  title={Focal loss for dense object detection},
  author={Lin, Tsung-Yi and Goyal, Priya and Girshick, Ross and He, Kaiming and Doll{\'a}r, Piotr},
  booktitle={Proceedings of the IEEE international conference on computer vision},
  pages={2980--2988},
  year={2017}
}

@inproceedings{woo2018cbam,
  title={Cbam: Convolutional block attention module},
  author={Woo, Sanghyun and Park, Jongchan and Lee, Joon-Young and Kweon, In So},
  booktitle={Proceedings of the European conference on computer vision (ECCV)},
  pages={3--19},
  year={2018}
}

@inproceedings{devlin2019bert,
    title = "{BERT}: Pre-training of Deep Bidirectional Transformers for Language Understanding",
    author = "Devlin, Jacob  and
      Chang, Ming-Wei  and
      Lee, Kenton  and
      Toutanova, Kristina",
    editor = "Burstein, Jill  and
      Doran, Christy  and
      Solorio, Thamar",
    booktitle = "Proceedings of the 2019 Conference of the North {A}merican Chapter of the Association for Computational Linguistics: Human Language Technologies, Volume 1 (Long and Short Papers)",
    month = jun,
    year = "2019",
    address = "Minneapolis, Minnesota",
    publisher = "Association for Computational Linguistics",
    doi = "10.18653/v1/N19-1423",
    pages = "4171--4186",
}

@misc{Security2025Antivirus,
  title = {{2025 Antivirus Trends, Statistics, and Market Report}},
  author = {{Security.org}},
  year = {2025},
    note = {Accessed: 2026-06-29},
  howpublished = {\url{https://www.security.org/antivirus/antivirus-consumer-report-annual/}}
}

@misc{Cybernews2025Antivirus,
  title = {{Antivirus Market Report 2026: Changes in User Behavior and New Trends}},
  author = {{Cybernews}},
  year = {2026},
    note = {Accessed: 2026-06-29},
  howpublished = {\url{https://cybernews.com/best-antivirus-software/antivirus-market-report/}}
}

@article{botacin2022antiviruses,
  title={Antiviruses under the microscope: A hands-on perspective},
  author={Botacin, Marcus and Domingues, Felipe Duarte and Ceschin, Fabr{\'\i}cio and Machnicki, Raphael and Alves, Marco Antonio Zanata and de Geus, Paulo L{\'\i}cio and Gr{\'e}gio, Andr{\'e}},
  journal={Computers \& Security},
  volume={112},
  pages={102500},
  year={2022},
  publisher={Elsevier}
}

@article{christodorescu2004testing,
  title={Testing malware detectors},
  author={Christodorescu, Mihai and Jha, Somesh},
  journal={ACM SIGSOFT Software Engineering Notes},
  volume={29},
  number={4},
  pages={34--44},
  year={2004},
  publisher={ACM New York, NY, USA}
}

@inproceedings{blackthorne2016avleak,
  title={$\{$AVLeak$\}$: fingerprinting antivirus emulators through $\{$Black-Box$\}$ testing},
  author={Blackthorne, Jeremy and Bulazel, Alexei and Fasano, Andrew and Biernat, Patrick and Yener, B{\"u}lent},
  booktitle={10th USENIX Workshop on Offensive Technologies (WOOT 16)},
  year={2016}
}

\end{document}